\documentclass[aps,prb,preprint,superscriptaddress,showpacs]{revtex4-1}
\usepackage{graphicx}
\usepackage{amsmath}
\usepackage{bm}

\begin{document}

\title{Optical properties of a two-nanospheroid cluster: analytical approach}

\author{D.V. Guzatov}
\email[]{guzatov@gmail.com} \affiliation{Yanka Kupala Grodno State
University, Ozheshko str. 22, 230023 Grodno, Belarus}
\author{V.V. Klimov}
\email[]{vklim@sci.lebedev.ru} \affiliation{P.N. Lebedev Physical
Institute, Russian Academy of Sciences, Leninsky pr. 53, 119991
Moscow, Russia}

\date{\today}

\begin{abstract}
Optical properties of a plasmonic nano-antenna made of two metallic
nanospheroids (prolate or oblate) are investigated analytically in
quasistatic approximation. It is shown that in clusters of two
nanospheroids, three types of plasmonic modes can be present. Two of
them can be effectively excited by a plane electromagnetic wave,
while the third one can be effectively excited only by a
nanolocalized light source (an atom, a molecule, a quantum dot)
placed in the gap between the nanoparticles. Analytical expressions
for absorption and scattering cross-sections, enhancement of a local
field, and radiative decay rate of a dipole source placed near such
a nano-antenna are presented and analyzed.
\end{abstract}


\maketitle

\section{\label{intro}Introduction}

Nowadays, quite a number of works are devoted to the study of
optical properties of single nanoparticles and their clusters.
Special attention is paid to metal nanoparticles with the help of
which it is possible to enhance electric fields at the frequencies
of localized plasmon resonances \cite{ref1,ref2,ref3}. On the basis
of this effect, a variety of possible applications are considered.
The most developed is use of large local fields near a rough surface
for increasing of surface-enhanced Raman scattering (SERS)
\cite{ref4}. Modification of fluorescence by means of nanoparticles
of different shapes is a basis for creation of nanobiosensors
\cite{ref5,ref6,ref7,ref8,ref9}, nano-antennas
\cite{ref10,ref11,ref12,ref13,ref14}, devices for decoding of DNA
structure \cite{ref15}, and etc.

At the present time, optical properties of single metallic
nanospheres and their clusters \cite{ref16,ref17,ref18,ref19,
ref20,ref21,ref22,ref23,ref24, ref25,ref26,ref27,ref28,ref29},
single nanospheroids \cite{ref30,ref31,ref32}, nanoellipsoids
\cite{ref33,ref34,ref35,ref36}, and some other nanobodies
\cite{ref2,ref3,ref37} are studied well enough from analytical point
of view. Many other nanoparticles' shapes and also nanoparticles
clusters are investigated only numerically. Unfortunately, numerical
simulations often do not allow understanding of physical nature of
interesting and complicated phenomena in this area. That is why any
analytical solutions will have principal importance.

In this work, we present the results of an analytical study of
optical properties of clusters of two metallic prolate or oblate
spheroidal nanoparticles. Such clusters are investigated both
experimentally and numerically and form a basis for various possible
applications including nanosensors, nano-antennas, and plasmon
waveguides \cite{ref38,ref39,ref40,ref41,ref42,ref43,ref44}. As far
as we know, there is no analytical investigation of optical
properties of two-nanospheroid clusters. That is caused by extremely
cumbersome mathematical calculations connected with use of general
spheroidal wave functions and general rotational-translational
addition theorems \cite{ref45,ref46,ref47,ref48}. However, in the
case of nanoparticles which sizes are less than the wavelength, i.e.
in a quasistatic case, one can neglect retardation effects. It
allows us to find an analytical solution for a cluster of two
spheroidal nanoparticles, placed in an arbitrary external field. The
geometry of the considered problem is shown in Fig.~\ref{fig1}. For
simplicity, we will consider that the cluster consists of two equal
spheroids made of a material with dielectric constant $\varepsilon$
and placed in vacuum. Generalization of our approach to unequal
spheroids is straightforward and can be easily done with the help of
the general theorem presented in the Appendix.

Main attention will be paid to a case of nearly touching and
strongly interacting spheroids since it is the case that seems to be
most interesting for applications because a substantial enhancement
of electric fields occurs there. The opposite case of weakly
interacting spheroids can be easily treated with approximation of
the spheroids by point dipoles with corresponding polarizabilities
\cite{ref3}.

For illustration of the analytical results obtained, we will
consider a case of two identical (prolate or oblate) nanospheroids
made of silver \cite{ref49}. We suppose that the largest size of the
nanospheroid is equal to 30 nm, and the aspect ratio of the spheroid
is taken to be equal to 0.6.

The rest of the paper is organized as follows. In the
Section~\ref{sect2}, free plasmon oscillations of a two-nanospheroid
cluster are investigated. The results of this section reveal
underlying physics and are necessary for interpretation of results
of the other sections. In the Section~\ref{sect3}, we will consider
optical properties of the two-nanospheroid cluster placed in the
field of a plane electromagnetic wave. Here, we will find absorption
and scattering cross-sections and the factor of local field
enhancement also. In the Section~\ref{sect4}, an object of
examination is optical properties of the two-nanospheroid cluster
placed in the field of a radiating atom or a molecule, and their
decay rates will be calculated. In the Appendix, derivation of the
translational addition theorem for spheroidal functions in a
quasistatic case is presented. It is the theorem that allows one to
find an analytical presentation of the solution for the
two-nanospheroid cluster.

\section{\label{sect2}Plasmon oscillations in a cluster of two nanospheroids}

It is well-known that all optical properties of nanoparticles can be
derived from their plasmonic spectra, that is, from related plasmon
eigenvalues $\varepsilon _{\nu} $ and eigenfunctions ${\mathbf
{e}}_{\nu}$, ${\mathbf {h}}_{\nu} $, that are solution of the
sourceless Maxwell equations \cite{ref50}:

\begin{eqnarray}
 {\textrm {rot}} {\mathbf {h}}_{\nu}  + i\left( {{\frac{{\omega}} {{v_{c}}} }}
\right)\tilde {\varepsilon} {\mathbf {e}}_{\nu}  = 0, \nonumber \\
 {\textrm {rot}} {\mathbf {e}}_{\nu}  - i\left( {{\frac{{\omega}} {{v_{c}}} }}
\right){\mathbf {h}}_{\nu}  = 0, \label{eq1}
\end{eqnarray}

\noindent where $\tilde {\varepsilon}  = \varepsilon _{\nu} $ inside
the nanoparticle and $\tilde {\varepsilon}  = 1$ outside it,
$\omega$ is frequency of electromagnetic oscillations and $v_{c}$ is
the speed of light in vacuum. As a result, electric field in the
presence of any nanoparticle can be presented in the following form
\cite{ref50}:

\begin{equation}
{\mathbf {E}}\left( {{\mathbf {r}}} \right) = {\mathbf {E}}_{0}
\left( {{\mathbf {r}}} \right) + {\sum\limits_{\nu}  {{\mathbf
{e}}_{\nu}  \left( {{\mathbf {r}}} \right)\left(
{{\frac{{\varepsilon \left( {\omega}  \right) - 1}}{{\varepsilon
_{\nu}  - \varepsilon \left( {\omega}  \right)}}}}
\right){\frac{{{\int\limits_{V} {\left( {{\mathbf {e}}_{\nu}  \left(
{{\mathbf {r}}} \right){\mathbf {E}}_{0} \left( {{\mathbf {r}}}
\right)} \right)dV}}} }{{{\int\limits_{V} {\left( {{\mathbf
{e}}_{\nu}  \left( {{\mathbf {r}}} \right){\mathbf {e}}_{\nu} \left(
{{\mathbf {r}}} \right)} \right)dV}}} }}}} , \label{eq2}
\end{equation}

\noindent where $\varepsilon \left( {\omega}  \right)$ describes the
dependence of dielectric permittivity of nanoparticle's specific
material on frequency $\omega$, ${\mathbf {E}}_{0}$ is the
excitation field, and $\nu$ is a vector index that defines specific
plasmonic mode. From Eq.~(\ref{eq2}), it is possible to find any
optical properties of a nanoparticle or a cluster of nanoparticles.
So, to understand very complicated optical properties of a
two-nanospheroid cluster, we should investigate plasmonic spectrum
of this system at first.

For studying plasmon oscillations and other optical properties of
clusters of two nanospheroids, it is enough to solve the quasistatic
equations:

\begin{equation}
 {\textrm {div}} \left( {\tilde {\varepsilon} {\mathbf {e}}_{\nu}}  \right) = 0, \quad {\textrm {rot}} {\mathbf {e}}_{\nu}  = 0, \label{eq3}
\end{equation}

\noindent that can be reduced to solution of Laplace equations by
substituting ${\mathbf {e}}_{\nu}  = - \nabla \varphi _{\nu}$:

\begin{eqnarray}
 \Delta \varphi _{\nu} ^{in} = 0, \quad && \textrm {inside the nanoparticle}, \nonumber \\
 \Delta \varphi _{\nu} ^{out} = 0, \quad && \textrm {outside the
 nanoparticle}, \nonumber \\
 {\left. {{\begin{array}{*{20}c}
 {{\left. {\varphi _{\nu} ^{in}}  \right|}_{S} = {\left. {\varphi _{\nu
}^{in}}  \right|}_{S} ,} \hfill \\
 {{\left. {\varepsilon _{\nu}  {\frac{{\partial \varphi _{\nu} ^{in}
}}{{\partial {\mathbf {n}}}}}} \right|}_{S} = {\left.
{{\frac{{\partial \varphi _{\nu} ^{out}}} {{\partial {\mathbf
{n}}}}}} \right|}_{S}}  \hfill
\\
\end{array}}}  \right\}} \quad &&{\begin{array}{*{20}c}
 {\textrm{at the surface}} \hfill \\
 {\textrm{of the nanoparticle}.} \hfill \\
\end{array}} \label{eq4}
 \end{eqnarray}

\noindent In Eq.~(\ref{eq4}), $\varphi _{\nu} ^{in}$, $\varphi
_{\nu} ^{out}$ are the potentials of plasmonic eigenfunctions inside
and outside the nanoparticle correspondingly, and ${\left.
{{\frac{{\partial \varphi _{\nu}} } {{\partial {\mathbf {n}}}}}}
\right|}_{S}$ denotes normal derivative at the nanoparticles'
surface $S$. The last equation in (\ref{eq4}) provides continuity of
normal components of electrical induction. Note that in this case
there is no need to find magnetic fields for description of
plasmonic oscillations.

The systems of equations obtained in such a way, have nontrivial
solutions only for some negative values of permittivity $\varepsilon
_{\nu}$ defining frequency of plasmon oscillations \cite{ref2,ref3}.
In the case of Drude theory, $\varepsilon \left( {\omega}  \right) =
1 - \omega _{pl}^{2} / \omega ^{2}$, frequency of plasmon
oscillations can be found from the expression

\begin{equation}
\omega _{\nu}  = {\frac{{\omega _{pl}}} {{\sqrt {1 - \varepsilon
_{\nu}}}}} < \omega _{pl} ,
\label{eq5}
\end{equation}

\noindent where $\omega _{pl} $ is bulk plasmon frequency of a metal
from which the nanoparticles are made. Our approach allows us to
investigate arbitrary spheroids, but for simplicity in the present
section we examine equations for plasmon oscillations in a cluster
of two identical metal nanospheroids.

In our case of a two-nanospheroid cluster, we will look for a
solution as follows. The total potential outside the spheroids will
be the sum of their partial potentials (we will omit the mode index
$\nu$ further) \cite{ref19,ref51}:

\begin{equation}
\varphi ^{out} = \varphi _{1}^{out} + \varphi _{2}^{out},
\label{eq6}
\end{equation}

\noindent while the potentials inside each nanospheroid shall be
denoted as $\varphi _{j}^{in}$ ($j = 1$, 2). To find $\varphi
^{out}$ and $\varphi _{1}^{in}$, $\varphi _{2}^{in}$, it is
naturally to use spheroidal coordinates. In the case of a prolate
nanospheroid, the relation between Cartesian and spheroidal
coordinates ($1 \le \xi < \infty $, $ - 1 \le \eta \le 1$, $0 \le
\phi \le 2\pi )$ is \cite{ref52}:

\begin{eqnarray}
 x &=& f\sqrt {\left( {\xi ^{2} - 1} \right)\left( {1 - \eta ^{2}} \right)}
\cos \phi , \nonumber \\
 y &=& f\sqrt {\left( {\xi ^{2} - 1} \right)\left( {1 - \eta ^{2}} \right)}
\sin \phi , \nonumber \\
 z &=& f\xi \eta ,
 \label{eq7}
\end{eqnarray}

\noindent where $f = \sqrt {c^{2} - a^{2}}$ is a half of the focal
distance in a prolate spheroid ($a < c$) which surface is set by the
equation $\left( {x^{2} + y^{2}} \right) / a^{2} + z^{2} / c^{2} =
1$.

In the case of an oblate spheroid ($a > c$), the relation between
Cartesian and spheroidal coordinates ($0 \le \xi < \infty $, $ - 1
\le \eta \le 1$, $0 \le \phi \le 2\pi$) has the following form
\cite{ref52}:

\begin{eqnarray}
 x &=& f\sqrt {\left( {\xi ^{2} + 1} \right)\left( {1 - \eta ^{2}} \right)}
\cos \phi , \nonumber \\
 y &=& f\sqrt {\left( {\xi ^{2} + 1} \right)\left( {1 - \eta ^{2}} \right)}
\sin \phi , \nonumber \\
 z &=& f\xi \eta ,
\label{eq8}
\end{eqnarray}

\noindent where $f = \sqrt {a^{2} - c^{2}}$ is a half of the focal
distance in the oblate spheroid. Let us note that this expression
can be obtained from Eq.~(\ref{eq7}) by substitution $\xi \to i\xi $
and $f \to - if$. Further, we will use this formal replacement since
it is fundamental and allows us to find a solution for oblate
spheroids if the solution for prolate spheroid geometry is known
\cite{ref51,ref52,ref53,ref54}.

\subsection{\label{subsect2a}Plasmon oscillations in a
cluster of two identical prolate nanospheroids}

\begin{figure}[there]
\includegraphics[width=9.0cm]{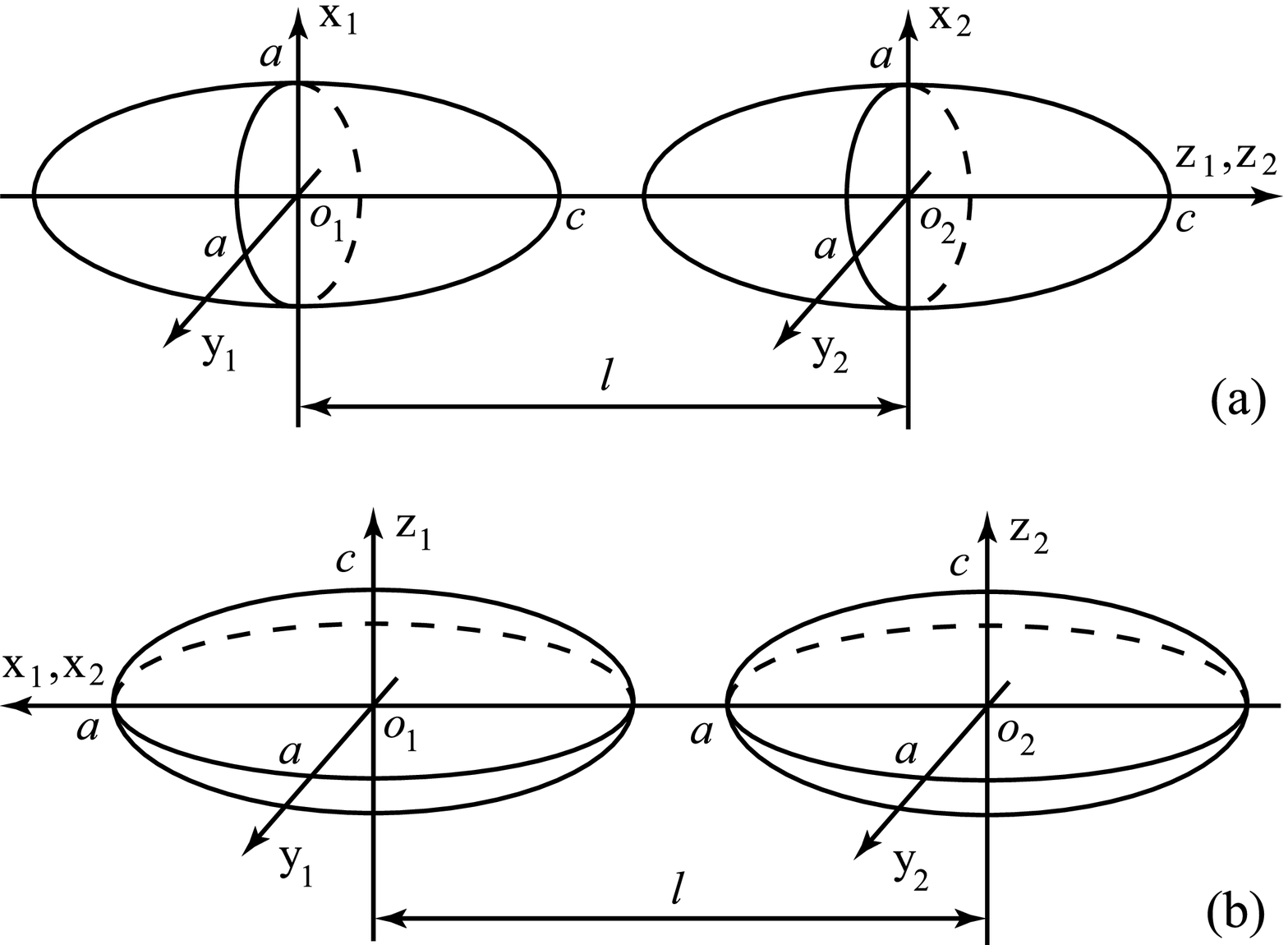}
\caption{\label{fig1} The geometry of a two-nanospheroid cluster.
(a) The case of prolate nanospheroids, (b) the case of oblate
nanospheroids.}
\end{figure}

To find plasmonic spectra of a two-nanospheroid cluster, it is
naturally to use two local systems of spheroidal coordinates ($\xi
_{j}$, $\eta _{j}$, $\phi _{j}$, $j = 1$, 2) the origins $o_{j}$ of
which are placed in the centers of corresponding nanospheroids (see
Fig.~\ref{fig1}(a)). Coordinates (and all other values) related to
the first or second nanospheroid will be denoted by the index ``1''
or ``2'' respectively. The potential inside the $j$-th nanospheroid
can be presented in the form \cite{ref55} ($j = 1$, 2):

\begin{equation}
\varphi _{j}^{in} = {\sum\limits_{n = 0}^{\infty} {{\sum\limits_{m =
0}^{n} {P_{n}^{m} \left( {\xi _{j}} \right)P_{n}^{m} \left( {\eta
_{j}} \right)\left( {A_{mn}^{\left( {j} \right)} \cos \left( {m\phi
_{j}}  \right) + B_{mn}^{\left( {j} \right)} \sin \left( {m\phi
_{j}}  \right)} \right)}}} }, \label{eq9}
\end{equation}

\noindent where $P_{n}^{m} \left( {\eta}  \right)$ is an associated
Legendre function \cite{ref56} defined in the region $ - 1 \le \eta
\le 1$, $P_{n}^{m} \left( {\xi}  \right)$ is an associated Legendre
function \cite{ref56} defined in a complex plane with the branch cut
from $ - \infty $ to +1. The partial potential outside the $j$-th
nanospheroid can be presented \cite{ref55} as ($j = 1$, 2)

\begin{equation}
\varphi _{j}^{out} = {\sum\limits_{n = 0}^{\infty} {{\sum\limits_{m
= 0}^{n} {Q_{n}^{m} \left( {\xi _{j}} \right)P_{n}^{m} \left( {\eta
_{j}} \right)\left( {C_{mn}^{\left( {j} \right)} \cos \left( {m\phi
_{j}}  \right) + D_{mn}^{\left( {j} \right)} \sin \left( {m\phi
_{j}}  \right)} \right)}}} } , \label{eq10}
\end{equation}

\noindent where $Q_{n}^{m} \left( {\xi}  \right)$ is an associated
Legendre function of the second kind \cite{ref56} defined in a
complex plane with the branch cut from $ - \infty $ to +1.

By construction, the potentials (\ref{eq9}) and (\ref{eq10}) are
solutions of the Laplace equation \cite{ref55}. So to find a
solution of Eq.~(\ref{eq4}), one should use only the boundary
conditions:

\begin{eqnarray}
 {\left. {\varphi _{1}^{in}}  \right|}_{\xi _{1} = \xi _{0}}  = {\left.
{\varphi ^{out}} \right|}_{\xi _{1} = \xi _{0}}  ,\quad {\left.
{\varepsilon {\frac{{\partial \varphi _{1}^{in}}} {{\partial \xi
_{1}}} }} \right|}_{\xi _{1} = \xi _{0}}  = {\left.
{{\frac{{\partial \varphi ^{out}}}{{\partial \xi
_{1}}} }} \right|}_{\xi _{1} = \xi _{0}}  , \nonumber \\
 {\left. {\varphi _{2}^{in}}  \right|}_{\xi _{2} = \xi _{0}}  = {\left.
{\varphi ^{out}} \right|}_{\xi _{2} = \xi _{0}}  ,\quad {\left.
{\varepsilon {\frac{{\partial \varphi _{2}^{in}}} {{\partial \xi
_{2}}} }} \right|}_{\xi _{2} = \xi _{0}}  = {\left.
{{\frac{{\partial \varphi ^{out}}}{{\partial \xi _{2}}} }}
\right|}_{\xi _{2} = \xi _{0}}, \label{eq11}
 \end{eqnarray}

\noindent where $\xi _{0} = c / \sqrt {c^{2} - a^{2}}  = c / f$ are
local radial coordinates defining surfaces of the nanospheroids,
$\varepsilon$ is permittivity of materials from which the
nanoparticles are made. To reduce the boundary conditions
(\ref{eq11}) to a system of linear equations, we apply the
translational addition theorem for wave functions of the prolate
nanospheroid (see the Appendix). In the case of two identical
coaxial nanospheroids, this theorem gives ($j$, $s = 1$, 2, $j \ne
s$, $\phi _{1} = \phi _{2}$)

\begin{equation}
Q_{n}^{m} \left( {\xi _{j}}  \right)P_{n}^{m} \left( {\eta _{j}}
\right) = {\sum\limits_{q = m}^{\infty} {S_{mqmn}^{\left( {j}
\right)} P_{q}^{m} \left( {\xi _{s}} \right)P_{q}^{m} \left( {\eta
_{s}}  \right)}} , \label{eq12}
\end{equation}

\noindent where the functions $S_{mqmn}^{\left( {1} \right)} =
S_{mqmn} \left( {f,f,l,0} \right)$ and $S_{mqmn}^{\left( {2}
\right)} = S_{mqmn} \left( {f,f,l,\pi}  \right)$ are defined in the
Appendix. Applying the boundary conditions (\ref{eq11}) and the
theorem (\ref{eq12}), one can obtain the following system of
equations ($n = 0$, 1, 2, \ldots ; $m = 0$, 1, 2, \ldots , $n$):

\begin{eqnarray}
 \left( {\varepsilon {\frac{{dP_{n}^{m} \left( {\xi _{0}}  \right)}}{{d\xi
_{0}}} }Q_{n}^{m} \left( {\xi _{0}}  \right) - P_{n}^{m} \left( {\xi
_{0}} \right){\frac{{dQ_{n}^{m} \left( {\xi _{0}}  \right)}}{{d\xi
_{0}}} }}
\right)C_{mn}^{\left( {1} \right)} \nonumber \\
 + \left( {\varepsilon - 1} \right)P_{n}^{m} \left( {\xi _{0}}
\right){\frac{{dP_{n}^{m} \left( {\xi _{0}}  \right)}}{{d\xi _{0}
}}}{\sum\limits_{q = m}^{\infty}  {S_{mnmq}^{\left( {0} \right)}
\left( { -
1} \right)^{m + q}C_{mq}^{\left( {2} \right)}}}   &=& 0, \nonumber \\
 \left( {\varepsilon {\frac{{dP_{n}^{m} \left( {\xi _{0}}  \right)}}{{d\xi
_{0}}} }Q_{n}^{m} \left( {\xi _{0}}  \right) - P_{n}^{m} \left( {\xi
_{0}} \right){\frac{{dQ_{n}^{m} \left( {\xi _{0}}  \right)}}{{d\xi
_{0}}} }}
\right)\left( { - 1} \right)^{m + n}C_{mn}^{\left( {2} \right)} \nonumber \\
 + \left( {\varepsilon - 1} \right)P_{n}^{m} \left( {\xi _{0}}
\right){\frac{{dP_{n}^{m} \left( {\xi _{0}}  \right)}}{{d\xi _{0}
}}}{\sum\limits_{q = m}^{\infty}  {S_{mnmq}^{\left( {0} \right)}
C_{mq}^{\left( {1} \right)}}}   &=& 0. \label{eq13}
 \end{eqnarray}

\noindent When deriving Eq.~(\ref{eq13}), we make use of the fact
that for identical nanospheroids $S_{mnmq}^{\left( {2} \right)} =
\left( { - 1} \right)^{n + q} S_{mnmq}^{\left( {1} \right)} $ (see
the Appendix) and take $S_{mnmq}^{\left( {0} \right)} = \left( { -
1} \right)^{m + n} S_{mnmq}^{\left( {1} \right)} $. The system of
equations for $D_{mn}^{\left( {j} \right)}$ is identical to
(\ref{eq13}) and gives no additional information for plasmonic
spectra of coaxial spheroids. So, we will not consider it further.

As it results from symmetry of the considered cluster and the system
(\ref{eq13}), there are two independent types of solutions
(plasmonic modes) with opposite parity. To select the first type of
the modes (symmetric in $z \to - z$ transformation), one should
choose $C_{mn}^{\left( {1} \right)} = \left( { - 1} \right)^{m + n}
C_{mn}^{\left( {2} \right)}$ in Eq.~(\ref{eq13}). As a result, we
shall obtain the following system of equations for the symmetric
modes:

\begin{eqnarray}
 \left( {\varepsilon {\frac{{dP_{n}^{m} \left( {\xi _{0}}  \right)}}{{d\xi
_{0}}} }Q_{n}^{m} \left( {\xi _{0}}  \right) - P_{n}^{m} \left( {\xi
_{0}} \right){\frac{{dQ_{n}^{m} \left( {\xi _{0}}  \right)}}{{d\xi
_{0}}} }}
\right)C_{mn}^{\left( {1} \right)} \nonumber \\
 + \left( {\varepsilon - 1} \right)P_{n}^{m} \left( {\xi _{0}}
\right){\frac{{dP_{n}^{m} \left( {\xi _{0}}  \right)}}{{d\xi _{0}
}}}{\sum\limits_{q = m}^{\infty}  {S_{mnmq}^{\left( {0} \right)}
C_{mq}^{\left( {1} \right)}}}   &=& 0. \label{eq14}
 \end{eqnarray}

To obtain the second type of the modes (antisymmetric in $z \to - z$
transformation), one should put $C_{mn}^{\left( {1} \right)} = -
\left( { - 1} \right)^{m + n}C_{mn}^{\left( {2} \right)}$ in
Eq.~(\ref{eq13}). As a result, we shall obtain the following system
of equations for the antisymmetric modes:

\begin{eqnarray}
 \left( {\varepsilon {\frac{{dP_{n}^{m} \left( {\xi _{0}}  \right)}}{{d\xi
_{0}}} }Q_{n}^{m} \left( {\xi _{0}}  \right) - P_{n}^{m} \left( {\xi
_{0}} \right){\frac{{dQ_{n}^{m} \left( {\xi _{0}}  \right)}}{{d\xi
_{0}}} }}
\right)C_{mn}^{\left( {1} \right)} \nonumber \\
 - \left( {\varepsilon - 1} \right)P_{n}^{m} \left( {\xi _{0}}
\right){\frac{{dP_{n}^{m} \left( {\xi _{0}}  \right)}}{{d\xi _{0}
}}}{\sum\limits_{q = m}^{\infty}  {S_{mnmq}^{\left( {0} \right)}
C_{mq}^{\left( {1} \right)}}}   &=& 0. \label{eq15}
 \end{eqnarray}

It is important to notice that separation of spectra into symmetric
and antisymmetric plasmon modes is possible only in the case when
there is a plane of symmetry. When $m$ is even, antisymmetric modes
have nonzero dipole moment and they are ``bright'' modes. On the
contrary, symmetric modes have zero dipole moment and are ``dark''
modes when $m$ is even. In the case of odd $m$, the ``bright'' and
``dark'' modes correspond to the symmetric and antisymmetric modes
respectively. One can expect that the antisymmetric mode $m = 0$
will have the largest polarizability and thus will be the
``brightest'' one for excitation of our cluster with a
longitudinally (along the z axis) polarized plane wave.

\begin{figure}[there]
\includegraphics[width=6.6cm]{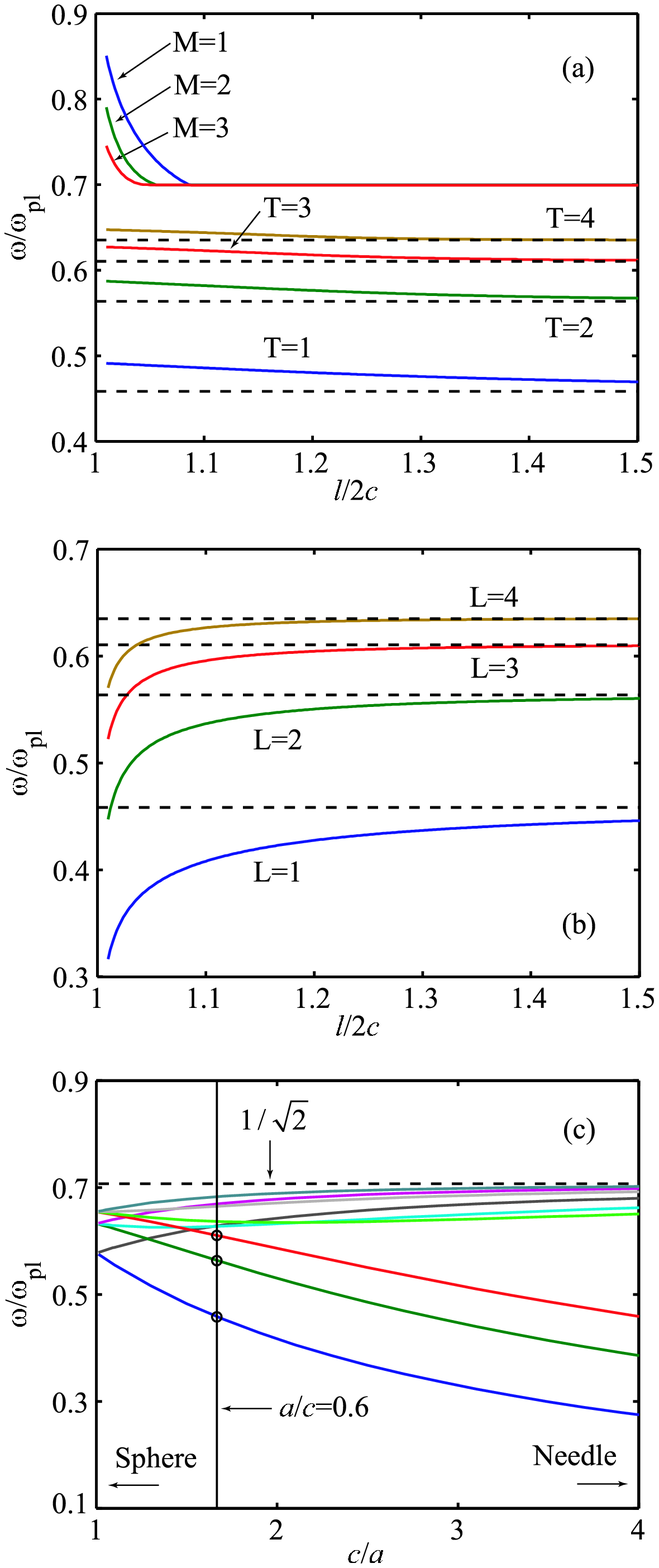}
\caption{\label{fig2} (Color online) Normalized frequencies of the
first three plasmonic oscillations in a cluster of two identical
prolate spheroidal nanoparticles as a function of the distance
between the nanospheroids' centers. The axis-symmetrical case ($m =
0$) is considered, and the aspect ratio of a single spheroid is $a/c
= 0.6$ ($\xi _{0} = 1 / \sqrt {1 - \left( {a / c} \right)^{2}} $).
(a) Symmetrical modes (eigenvalues of Eq.~(\ref{eq14})), (b)
antisymmetric modes (eigenvalues of Eq.~(\ref{eq15})). Dashed lines
show plasmon frequencies of a single prolate nanospheroid
${\frac{{\omega}} {{\omega _{pl}}} } = {\left\{ {{\frac{{dP_{n}^{m}
\left( {\xi _{0}}  \right)}}{{d\xi _{0}}} }Q_{n}^{m} \left( {\xi
_{0}}  \right) / \left( {{\frac{{dP_{n}^{m} \left( {\xi _{0}}
\right)}}{{d\xi _{0} }}}Q_{n}^{m} \left( {\xi _{0}}  \right) -
P_{n}^{m} \left( {\xi _{0}} \right){\frac{{dQ_{n}^{m} \left( {\xi
_{0}}  \right)}}{{d\xi _{0}}} }} \right)} \right\}}^{1 / 2}$. The
figure (c) shows plasmon frequencies of the single prolate
nanospheroid as a function of the inverse aspect ratio $c/a$. The
vertical line corresponds to $a/c = 0.6$ and allows to select
asymptotic values for the figures (a) and (b).}
\end{figure}

To study plasmon oscillations in clusters of two prolate spheroidal
nanoparticles, we have solved the eigenvalue problems (\ref{eq14})
and (\ref{eq15}) numerically. In Fig.~\ref{fig2}, normalized plasmon
frequency $\omega / \omega _{pl}$ of a cluster of two prolate
nanospheroids (see Fig.~\ref{fig1}(a) for the geometry),
corresponding to the first three plasmon modes, is shown as a
function of normalized distances $l/2c$ between the nanoparticles'
centers. Eigenvalues $\varepsilon $ have been obtained as a
nontrivial solution of the equations systems (\ref{eq14}) and
(\ref{eq15}) in the case of an axis-symmetric problem ($m = 0$).
Then, the found solutions have been substituted into Eq.~(\ref{eq5})
to obtain plasmon oscillations frequency.

In Fig.~\ref{fig2}, one can observe that plasmon frequencies of a
cluster of two prolate nanospheroids tend to plasmon frequencies of
a single nanospheroid (see Fig.~\ref{fig2}(c)) if the distances
between the nanospheroids are large enough. When width of the gap
between the nanospheroids tends to zero, solutions of the equations
(\ref{eq14}) and (\ref{eq15}) behave very differently. For symmetric
modes (Fig.~\ref{fig2}(a)), there are two branches: T-modes and
M-modes. Modes of ``T'' type can be obtained by the method of
hybridization of plasmon modes of a single prolate nanospheroid,
analogously to the case of a cluster of two spherical nanoparticles
\cite{ref25}. When width of the gap between the nanoparticles is
decreasing to zero, normalized plasmonic frequencies of T-modes tend
to various values in the range from 0 to $1 / \sqrt {2}$, on the
analogy with a two-sphere cluster \cite{ref20,ref57}. T-modes with
higher indices (not shown for clarity) will concentrate near $\omega
/ \omega _{pl} = 1 / \sqrt {2}$. In Fig.~\ref{fig2}(a), one can also
see that at very short distances between the nanospheroids ($l / 2c
< 1.1$), a new type of plasmonic modes (M-modes) appears. M-modes
are characterized by strong spatial localization in the gap between
the nanoparticles. As a result, they can be effectively excited only
by a strongly nonuniform electric field of the molecule or the
quantum dot. Values of plasmonic frequencies of these modes lie in
the range $\omega _{pl} / \sqrt {2} < \omega < \omega _{pl}$. As the
gap width decreases to zero, plasmon frequency of M-modes tends to
bulk plasmon frequency $\omega _{pl}$.

In Fig.~\ref{fig2}(b), nontrivial solutions of the equations system
(\ref{eq15}) for the antisymmetric potential in an axial-symmetric
case ($m = 0$), are shown. By analogy with a two-sphere cluster, we
will call these modes L-modes (longitudinal) because they are
"bright" only for longitudinal excitation. These modes can be
described by the hybridization method of plasmon oscillations of
single nanospheroids forming the considered cluster. As width of the
gap between prolate nanospheroids decreases to zero, normalized
plasmon frequencies of these modes tend to zero as it also takes
place in the case of spherical nanoparticles \cite{ref20,ref57}.
Plasmonic frequencies of L-modes of higher orders (not shown) tend
to $\omega _{pl} / \sqrt {2}$, and concentration of infinite number
of L-modes occurs near this value.

\begin{figure}[there]
\includegraphics[width=11.0cm]{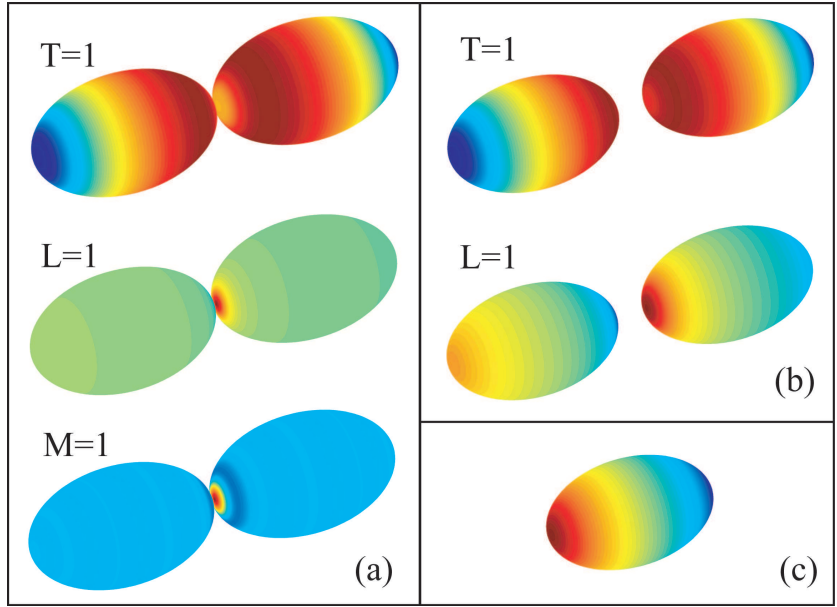}
\caption{\label{fig3} (Color online) Distribution of a surface
charge (a.u.) of the lowest plasmon mode in a cluster of two
identical prolate spheroidal nanoparticles according to solution of
the equations (\ref{eq14}) and (\ref{eq15}). The axial-symmetrical
case ($m = 0$) is considered, and the aspect ratio of a single
spheroid is $a/c = 0.6$. The distance between the nanospheroids'
centers is $l/2c = 1.03$ (a), $l/2c = 1.2$ (b). Distribution of a
surface charge in a single spheroid is shown in panel (c). The red
color corresponds to the positive charge, and blue - to the negative
one.}
\end{figure}

In Fig.~\ref{fig3}, distribution of a surface charge of plasmonic
modes of the lowest order in clusters of two identical prolate
nanospheroids, is shown. It is seen in this figure that the T- and
M-modes have symmetric distribution of the surface charge in
contrast to the antisymmetric L=1 mode. This behavior, of course, is
in agreement with symmetry of the equations (\ref{eq14}) and
(\ref{eq15}). Another interesting feature is that the surface charge
of T-modes is distributed over the surface of the all nanoparticles
for any distances between them, while for M- and L-modes it is
concentrated near the gap between the nanospheroids if the distance
between them is sufficiently small. It is interesting to note also
that the surface charge of M-modes is more concentrated in
comparison with that of L-modes. Indeed, due to an electroneutrality
requirement, the total surface charge on each nanospheroid should be
equal to zero. Here, both positive and negative charges of M-modes
are localized near the gap between the nanoparticles so that on the
rest of nanoparticles the charge is almost equal to zero, as it is
well seen in Fig.~\ref{fig3}(a). At the same time, in the case of
L-modes for each of nanospheroids near the gap a charge of only one
sign is concentrated, and a charge of the opposite sign is
distributed with small magnitude over the remaining surface of the
nanoparticles. Therefore, strictly speaking, the surface charge in
an L-mode is distributed over the whole surface of the cluster
nanoparticles though it is not clearly seen at small distances
between the nanoparticles (see Fig.~\ref{fig3}(a)). As the distance
increases, the charge distribution changes in the cluster: it
spreads over the nanoparticles' surface, tending in the limit to a
distribution corresponding to single prolate nanospheroids (see
Fig.~\ref{fig3}(c)).

\subsection{\label{subsect2b}Plasmon oscillations in a cluster of
two identical oblate nanospheroids}

In this geometry, one should also use local systems of coordinates
($\xi _{j}$, $\eta _{j}$, $\phi _{j}$, $j = 1$, 2) which are
connected with each nanospheroid and have origins $o_{j}$ in their
centers (see Fig.~\ref{fig1}(b)). Now, the electric potential inside
the $j$-th nanospheroid can be presented in the form ($j = 1$, 2):

\begin{equation}
\varphi _{j}^{in} = {\sum\limits_{n = 0}^{\infty} {{\sum\limits_{m =
0}^{n} {P_{n}^{m} \left( {i\xi _{j}} \right)P_{n}^{m} \left( {\eta
_{j}} \right)\left( {A_{mn}^{\left( {j} \right)} \cos \left( {m\phi
_{j}}  \right) + B_{mn}^{\left( {j} \right)} \sin \left( {m\phi
_{j}}  \right)} \right)}}} }, \label{eq16}
\end{equation}

\noindent and the partial potential outside the $j$-th oblate
nanospheroid will look like ($j = 1$, 2):

\begin{equation}
\varphi _{j}^{out} = {\sum\limits_{n = 0}^{\infty} {{\sum\limits_{m
= 0}^{n} {Q_{n}^{m} \left( {i\xi _{j}} \right)P_{n}^{m} \left( {\eta
_{j}} \right)\left( {C_{mn}^{\left( {j} \right)} \cos \left( {m\phi
_{j}}  \right) + D_{mn}^{\left( {j} \right)} \sin \left( {m\phi
_{j}}  \right)} \right)}}} }. \label{eq17}
\end{equation}

\noindent The total potential outside the nanospheroids will be
expressed by Eq.~(\ref{eq6}). As boundary conditions for the
potential, Eq.~(\ref{eq11}), where $\xi _{0} = c / \sqrt {a^{2} -
c^{2}} = c / f$, is used. In the case of oblate nanospheroids, the
addition translation theorem (see the Appendix) has the following
form ($j$, $s = 1$, 2, $j \ne s$):

\begin{equation}
 Q_{n}^{m} \left( {i\xi _{j}}  \right)P_{n}^{m} \left( {\eta _{j}}
\right){\left\{ {{\begin{array}{*{20}c}
 {\cos \left( {m\phi _{j}}  \right)} \hfill \\
 {\sin \left( {m\phi _{j}}  \right)} \hfill \\
\end{array}}}  \right.}
 = {\sum\limits_{q = 0}^{\infty}  {{\sum\limits_{p = 0}^{q} {P_{q}^{p}
\left( {i\xi _{s}}  \right)P_{q}^{p} \left( {\eta _{s}}
\right){\left\{ {{\begin{array}{*{20}c}
 {M_{pqmn}^{\left( {j} \right)} \cos \left( {p\phi _{s}}  \right)} \hfill \\
 {N_{pqmn}^{\left( {j} \right)} \sin \left( {p\phi _{s}}  \right)} \hfill \\
\end{array}}}  \right.}}}} } , \label{eq18}
 \end{equation}

\noindent where the functions $M_{pqmn}^{\left( {1} \right)} =
M_{pqmn} \left( { - if , - if ,l,\pi}  \right)$, $N_{pqmn}^{\left(
{1} \right)} = N_{pqmn} \left( { - if , - if ,l,\pi}  \right)$ and
$M_{pqmn}^{\left( {2} \right)} = M_{pqmn} \left( { - if, - if,l,0}
\right)$, $N_{pqmn}^{\left( {2} \right)} = N_{pqmn} \left( { - if, -
if,l,0} \right)$ are defined in the Appendix. Now, substituting
Eqs.~(\ref{eq16}) and (\ref{eq17}) into Eq.~(\ref{eq11}) and making
use of the addition translation theorem (\ref{eq18}), we shall
obtain the following system of equations ($n = 0$, 1, 2, \ldots ; $m
= 0$, 1, 2, \ldots , $n$):

\begin{eqnarray}
 \left( {\varepsilon {\frac{{dP_{n}^{m} \left( {i\xi _{0}}  \right)}}{{d\xi
_{0}}} }Q_{n}^{m} \left( {i\xi _{0}}  \right) - P_{n}^{m} \left(
{i\xi _{0} } \right){\frac{{dQ_{n}^{m} \left( {i\xi _{0}}
\right)}}{{d\xi _{0}}} }}
\right)C_{mn}^{\left( {1} \right)} \nonumber \\
 + \left( {\varepsilon - 1} \right)P_{n}^{m} \left( {i\xi _{0}}
\right){\frac{{dP_{n}^{m} \left( {i\xi _{0}}  \right)}}{{d\xi _{0}
}}}{\sum\limits_{q = 0}^{\infty}  {{\sum\limits_{p = 0}^{q}
{M_{mnpq}^{\left( {0} \right)} \left( { - 1}
\right)^{p}C_{pq}^{\left( {2}
\right)}}} } }  &=& 0, \nonumber \\
 \left( {\varepsilon {\frac{{dP_{n}^{m} \left( {i\xi _{0}}  \right)}}{{d\xi
_{0}}} }Q_{n}^{m} \left( {i\xi _{0}}  \right) - P_{n}^{m} \left(
{i\xi _{0} } \right){\frac{{dQ_{n}^{m} \left( {i\xi _{0}}
\right)}}{{d\xi _{0}}} }}
\right)\left( { - 1} \right)^{m}C_{mn}^{\left( {2} \right)} \nonumber \\
 + \left( {\varepsilon - 1} \right)P_{n}^{m} \left( {i\xi _{0}}
\right){\frac{{dP_{n}^{m} \left( {i\xi _{0}}  \right)}}{{d\xi _{0}
}}}{\sum\limits_{q = 0}^{\infty}  {{\sum\limits_{p = 0}^{q}
{M_{mnpq}^{\left( {0} \right)} C_{pq}^{\left( {1} \right)}}} } }
&=& 0. \label{eq19}
 \end{eqnarray}

\noindent By deriving Eq.~(\ref{eq19}), we take into account that
$M_{mnpq}^{\left( {2} \right)} = \left( { - 1} \right)^{m + p}
M_{mnpq}^{\left( {1} \right)}$ (see the Appendix) and denote
$M_{mnpq}^{\left( {0} \right)} = \left( { - 1} \right)^{m}
M_{mnpq}^{\left( {1} \right)}$. The system of equations for
$D_{mn}^{\left( {j} \right)}$ is analogous to Eq.~(\ref{eq19}), and
we will not analyze it here.

Due to symmetry of a cluster of two identical oblate nanospheroids,
there are two types of plasmon oscillation: symmetric and
antisymmetric relatively the symmetry plane. To select the first
(symmetric in $x \to - x$ transformation) type, we take
$C_{mn}^{\left( {1} \right)} = \left( { - 1} \right)^{m}
C_{mn}^{\left( {2} \right)} $. As a result, we shall obtain the
system of equations:

\begin{eqnarray}
 \left( {\varepsilon {\frac{{dP_{n}^{m} \left( {i\xi _{0}}  \right)}}{{d\xi
_{0}}} }Q_{n}^{m} \left( {i\xi _{0}}  \right) - P_{n}^{m} \left(
{i\xi _{0} } \right){\frac{{dQ_{n}^{m} \left( {i\xi _{0}}
\right)}}{{d\xi _{0}}} }}
\right)C_{mn}^{\left( {1} \right)} \nonumber \\
 + \left( {\varepsilon - 1} \right)P_{n}^{m} \left( {i\xi _{0}}
\right){\frac{{dP_{n}^{m} \left( {i\xi _{0}}  \right)}}{{d\xi _{0}
}}}{\sum\limits_{q = 0}^{\infty}  {{\sum\limits_{p = 0}^{q}
{M_{mnpq}^{\left( {0} \right)} C_{pq}^{\left( {1} \right)}}} } }
&=& 0. \label{eq20}
 \end{eqnarray}

To select the second (antisymmetric in $x \to - x$ transformation)
type of plasmon modes in Eq.~(\ref{eq19}), we take $C_{mn}^{\left(
{1} \right)} = - \left( { - 1} \right)^{m}C_{mn}^{\left( {2}
\right)}$. As a result, we shall have

\begin{eqnarray}
 \left( {\varepsilon {\frac{{dP_{n}^{m} \left( {i\xi _{0}}  \right)}}{{d\xi
_{0}}} }Q_{n}^{m} \left( {i\xi _{0}}  \right) - P_{n}^{m} \left(
{i\xi _{0} } \right){\frac{{dQ_{n}^{m} \left( {i\xi _{0}}
\right)}}{{d\xi _{0}}} }}
\right)C_{mn}^{\left( {1} \right)} \nonumber \\
 - \left( {\varepsilon - 1} \right)P_{n}^{m} \left( {i\xi _{0}}
\right){\frac{{dP_{n}^{m} \left( {i\xi _{0}}  \right)}}{{d\xi _{0}
}}}{\sum\limits_{q = 0}^{\infty}  {{\sum\limits_{p = 0}^{q}
{M_{mnpq}^{\left( {0} \right)} C_{pq}^{\left( {1} \right)}}} } }
&=& 0. \label{eq21}
 \end{eqnarray}

\begin{figure}[there]
\includegraphics[width=6.6cm]{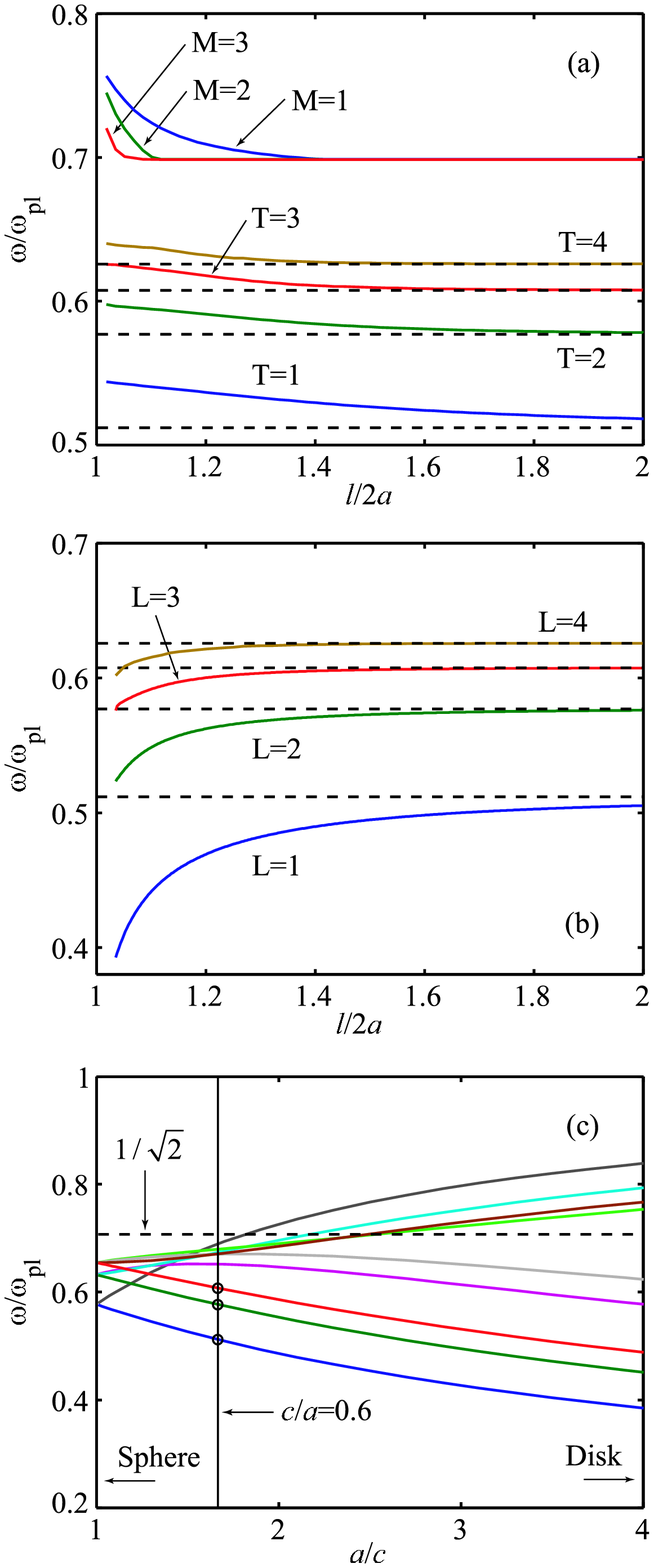}
\caption{\label{fig4} (Color online) Normalized plasmon frequencies
of the first three modes in a cluster of two identical oblate
nanospheroids as a function of the distance between their centers.
(a) Symmetrical modes (the eigenvalues of Eq.~(\ref{eq20})), (b)
antisymmetric modes (the eigenvalues of Eq.~(\ref{eq21})). By dashed
lines, plasmon frequencies of a single oblate nanospheroid
${\frac{{\omega}} {{\omega _{pl}}} } = {\left\{ {{\frac{{dP_{n}^{m}
\left( {i\xi _{0}} \right)}}{{d\xi _{0}}} }Q_{n}^{m} \left( {i\xi
_{0}}  \right) / \left( {{\frac{{dP_{n}^{m} \left( {i\xi _{0}}
\right)}}{{d\xi _{0} }}}Q_{n}^{m} \left( {i\xi _{0}}  \right) -
P_{n}^{m} \left( {i\xi _{0}} \right){\frac{{dQ_{n}^{m} \left( {i\xi
_{0}}  \right)}}{{d\xi _{0}}} }} \right)} \right\}}^{1 / 2}$ are
shown. The spect ratio of a single oblate nanospheroids is $c/a =
0.6$ ($\xi _{0} = 1 / \sqrt {\left( {a / c} \right)^{2} - 1} $). The
figure (c) shows plasmon frequencies of a single oblate nanospheroid
as a function of the inverse aspect ratio $a/c$. The vertical line
corresponds to $c/a = 0.6$ and allows to select asymptotic values
for the figures (a) and (b).}
\end{figure}

In Fig.~\ref{fig4}, the dependence of normalized plasmon frequencies
$\omega / \omega _{pl} = 1 / \sqrt {1 - \varepsilon}$ of a cluster
of two identical oblate nanospheroids on normalized distances $l/2a$
between the nanoparticles' centers, is shown for the first three
plasmon modes. Eigenvalues $\varepsilon$ were obtained as a solution
of the equations systems (\ref{eq20}) and (\ref{eq21}).

One can see in Fig.~\ref{fig4} that in clusters of two oblate
nanospheroids, modes of ``T'', ``M'', and ``L'' types, that are
analogous to T-, M- and L-modes of a cluster made of two prolate
spheroids (see Fig.~\ref{fig2}), can exist. The T- and M-modes are
the solutions of the system (\ref{eq20}), while the L-modes are the
solution of the system (\ref{eq21}). T- and L-modes can be derived
by the method of hybridization of plasmonic modes of two oblate
nanospheroids and their plasmonic frequencies are lying in the range
$0 < \omega < \omega _{pl} / \sqrt {2}$. An infinite number of
plasmonic frequencies of higher L-and T-modes lie near $\omega _{pl}
/ \sqrt {2}$. When width of the gap decreases to 0, the ratio
$\omega / \omega _{pl} $ for T-modes tends to various values in the
range from 0 to $1 / \sqrt {2} $, while plasmonic frequencies of
L-modes approach zero by analogy with L-modes in a cluster of two
spherical nanoparticles \cite{ref20,ref57}. Plasmonic frequencies of
strongly localized M-modes (Fig.~\ref{fig4}(a)) lie in the range
$\omega _{pl} / \sqrt {2} < \omega < \omega _{pl}$, as it happens in
a cluster of two spherical nanoparticles \cite{ref20,ref57}. As
width of the gap between oblate nanospheroids decreases to zero,
plasmon frequencies of M-modes tend to bulk plasmon frequency
$\omega _{pl}$ analogously to the case of a two-sphere cluster
\cite{ref20,ref57}. For large distances between the spheroids,
M-modes disappear, and plasmon frequencies of L- and T-modes of a
cluster of two oblate nanospheroids tend to plasmonic frequencies of
a single spheroid (see. Fig.~\ref{fig4}(c)) and can be found by
means of a self-consistent model with approximation of spheroids by
anisotropic point dipoles.

\begin{figure}[there]
\includegraphics[width=11.0cm]{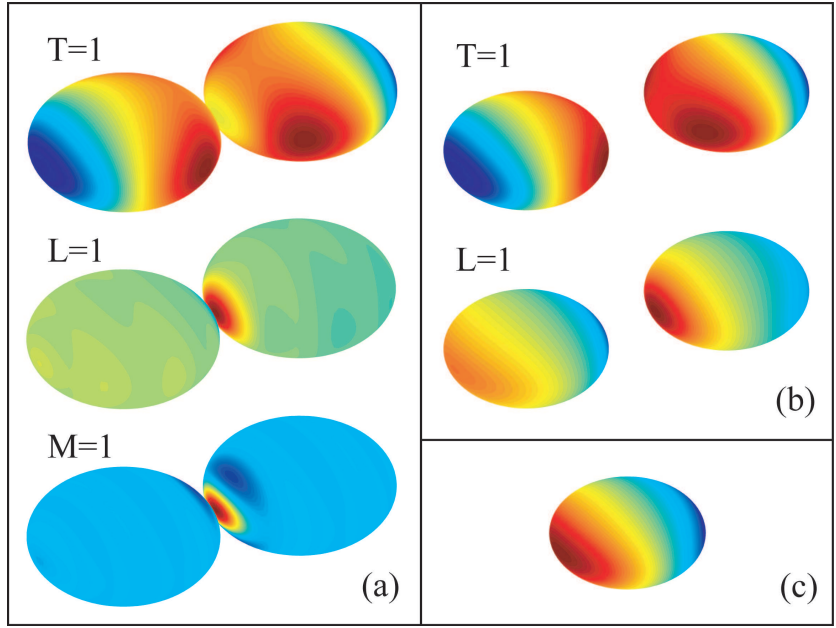}
\caption{\label{fig5} (Color online) Distribution of a surface
charge (a.u.) of the lowest plasmon modes in a cluster of two
identical oblate spheroidal nanoparticles according to solution of
the equations (\ref{eq20}) and (\ref{eq21}). The aspect ratio is
$c/a = 0.6$. The distance between the nanospheroids' centers $l/2a =
1.05$ (a), $l/2a = 1.4$ (b). Distribution of a surface charge in a
single spheroid is shown in panel (c). The red color corresponds to
the positive charge, and blue - to the negative one.}
\end{figure}

In Fig.~\ref{fig5}, the distribution of a surface charge of plasmon
modes of lower orders in a cluster of two identical oblate
nanospheroids is shown. One can see in this figure that the charge
distribution is symmetric in T- and M-modes, while in L-mode it is
antisymmetric in agreement with the definition of these modes. For
small distances between nanospheroids, charges in M- and L-modes are
strongly localized near the gap. On the contrary, when the distance
between the spheroids increases, the charge distribution tends to
symmetric or antisymmetric combination of a surface charge in a
single oblate nanospheroid (see Fig.~\ref{fig5}(c)).

Thus, in a cluster of two oblate or prolate spheroidal
nanoparticles, fundamental symmetric and antisymmetric plasmon modes
of ``T'', ``M'', and ``L'' types can be excited, and it is these
modes that define all optical properties of a two-nanospheroid
cluster.

\section{\label{sect3}A cluster of two metal nanospheroids in the field of a plane
electromagnetic wave}

In this section, we will consider a two-spheroid cluster in a
uniform electric field with the potential

\begin{equation}
\varphi _{0} = - E_{0x} x - E_{0y} y - E_{0z} z, \label{eq22}
\end{equation}

\noindent where the time factor $e^{ - i\omega t}$ is omitted. This
case corresponds to a plane wave incidence and is important for
transformation of far fields into near fields, for enhancement of
electric fields, and for effective excitation of atoms and
molecules. Here again, we restrict ourselves to two special cases of
nanospheroids' shape and position which are most interesting for
applications. In the first case, two prolate nanospheroids have a
general axis of rotation z, and axes x and y are parallel
(Fig.~\ref{fig1}(a)). In the second case, two oblate nanospheroids
have parallel axes of rotation z and general axis x, i.e., the
centers of the nanoparticles are located in the plane that is
perpendicular to the axes of rotation (Fig.~\ref{fig1}(b)). The
distance between the nanospheroids' centers is more than a half-sum
of the lengths between their foci (the nanoparticles are not
overlapping).

\subsection{\label{sect3a}A cluster of two prolate nanospheroids}

Here, we also will use local systems of spheroidal coordinates, the
origins of which are placed in the nanospheroids' centers (see
Fig.~\ref{fig1}(a)). The potential inside the $j$-th nanospheroid
again can be presented as a series in spheroidal harmonics
(\ref{eq9}), while the potential outside the nanospheroids now
should be presented in the form:

\begin{equation}
\varphi ^{out} = \varphi _{1}^{out} + \varphi _{2}^{out} + \varphi
_{0} , \label{eq23}
\end{equation}

\noindent where $\varphi _{1}^{out}$, $\varphi _{2}^{out}$ are
contributions from the first and second nanospheroids (see
Eq.~(\ref{eq10})), and $\varphi _{0}$ is the potential of the
external electric field (\ref{eq22}).

Electric potential of the incident plane wave (\ref{eq22}) in local
coordinates of the $j$-th ($j = 1$, 2) prolate nanospheroids, looks
like:

\begin{eqnarray}
 \varphi _{0}^{\left( {j} \right)} = &&f\left( {E_{0x} \cos \left( {\phi _{j}
} \right) + E_{0y} \sin \left( {\phi _{j}}  \right)}
\right)P_{1}^{1} \left(
{\xi _{j}}  \right)P_{1}^{1} \left( {\eta _{j}}  \right) \nonumber \\
 &&- fE_{0z} P_{1} \left( {\xi _{j}}  \right)P_{1} \left( {\eta _{j}}  \right)
+ \left( { - 1} \right)^{j + 1}E_{0z} {\frac{{l}}{{2}}},
\label{eq24}
 \end{eqnarray}

To find electric potentials and electric fields inside and outside
the spheroids, one should use the boundary conditions for the
potential:

\begin{eqnarray}
 {\left. {\varphi _{1}^{in}}  \right|}_{\xi _{1} = \xi _{0}}  &=& {\left.
{\left( {\varphi _{1}^{out} + \varphi _{2}^{out} + \varphi
_{0}^{\left( {1} \right)}}  \right)} \right|}_{\xi _{1} = \xi _{0}}
,\nonumber \\
{\left. {\varepsilon {\frac{{\partial \varphi _{1}^{in}}} {{\partial
\xi _{1}}} }} \right|}_{\xi _{1} = \xi _{0}} &=& {\left.
{{\frac{{\partial \left( {\varphi _{1}^{out} + \varphi _{2}^{out} +
\varphi _{0}^{\left( {1} \right)}}
\right)}}{{\partial \xi _{1}}} }} \right|}_{\xi _{1} = \xi _{0}}  , \nonumber \\
 {\left. {\varphi _{2}^{in}}  \right|}_{\xi _{2} = \xi _{0}}  &=& {\left.
{\left( {\varphi _{1}^{out} + \varphi _{2}^{out} + \varphi
_{0}^{\left( {2} \right)}}  \right)} \right|}_{\xi _{2} = \xi _{0}}
, \nonumber \\
{\left. {\varepsilon {\frac{{\partial \varphi _{2}^{in}}} {{\partial
\xi _{2}}} }} \right|}_{\xi _{2} = \xi _{0}} &=& {\left.
{{\frac{{\partial \left( {\varphi _{1}^{out} + \varphi _{2}^{out} +
\varphi _{0}^{\left( {2} \right)}} \right)}}{{\partial \xi _{2}}} }}
\right|}_{\xi _{2} = \xi _{0}}  , \label{eq25}
\end{eqnarray}

\noindent where $\xi _{0} = c / \sqrt {c^{2} - a^{2}}  = c / f$ is
local radial coordinate that defines surfaces of the nanoparticles,
$\varepsilon$ is permittivity of the nanoparticle's material. Making
use of the boundary conditions (\ref{eq25}) and the addition
translation theorem (\ref{eq12}), one can obtain the following
system of equations for the coefficients $C_{mn}^{\left( {1}
\right)}$ , $C_{mn}^{\left( {2} \right)}$, $D_{mn}^{\left( {1}
\right)}$ and $D_{mn}^{\left( {2} \right)}$ that define the outside
field (see Eq.~(\ref{eq10}) and Eq.~(\ref{eq23})) ($n = 0$, 1, 2,
\ldots ; $m = 0$, 1, 2, \ldots , $n$):

\begin{eqnarray}
\left( {\varepsilon {\frac{{dP_{n}^{m} \left( {\xi _{0}}
\right)}}{{d\xi _{0}}} }Q_{n}^{m} \left( {\xi _{0}}  \right) -
P_{n}^{m} \left( {\xi _{0}} \right){\frac{{dQ_{n}^{m} \left( {\xi
_{0}} \right)}}{{d\xi _{0}}} }}
\right)C_{mn}^{\left( {1} \right)} && \nonumber \\
 + \left( {\varepsilon - 1} \right)P_{n}^{m} \left( {\xi _{0}}
\right){\frac{{dP_{n}^{m} \left( {\xi _{0}}  \right)}}{{d\xi _{0}
}}}{\sum\limits_{q = m}^{\infty}  {S_{mnmq}^{\left( {2} \right)}
C_{mq}^{\left( {2} \right)}}} && = a_{mn}^{\left( {1} \right)}
\left( {\varepsilon - 1} \right)P_{n}^{m} \left( {\xi _{0}}
\right){\frac{{dP_{n}^{m} \left( {\xi _{0}}  \right)}}{{d\xi _{0}}} }, \nonumber \\
\left( {\varepsilon {\frac{{dP_{n}^{m} \left( {\xi _{0}}
\right)}}{{d\xi _{0}}} }Q_{n}^{m} \left( {\xi _{0}}  \right) -
P_{n}^{m} \left( {\xi _{0}} \right){\frac{{dQ_{n}^{m} \left( {\xi
_{0}}  \right)}}{{d\xi _{02}}} }}
\right)C_{mn}^{\left( {2} \right)}&& \nonumber \\
 + \left( {\varepsilon - 1} \right)P_{n}^{m} \left( {\xi _{0}}
\right){\frac{{dP_{n}^{m} \left( {\xi _{0}}  \right)}}{{d\xi _{0}
}}}{\sum\limits_{q = m}^{\infty}  {S_{mnmq}^{\left( {1} \right)}
C_{mq}^{\left( {1} \right)}}} &&= a_{mn}^{\left( {2} \right)} \left(
{\varepsilon - 1} \right)P_{n}^{m} \left( {\xi _{0}}
\right){\frac{{dP_{n}^{m} \left( {\xi _{0}}  \right)}}{{d\xi _{0}}}
}, \nonumber \\ \label{eq26}
\end{eqnarray}

\noindent and ($n = 1$, 2, 3, \ldots ; $m = 1$, 2, 3, \ldots , $n$)

\begin{eqnarray}
 \left( {\varepsilon {\frac{{dP_{n}^{m} \left( {\xi _{0}}  \right)}}{{d\xi
_{0}}} }Q_{n}^{m} \left( {\xi _{0}}  \right) - P_{n}^{m} \left( {\xi
_{01}} \right){\frac{{dQ_{n}^{m} \left( {\xi _{0}}  \right)}}{{d\xi
_{0}}} }}
\right)D_{mn}^{\left( {1} \right)} && \nonumber \\
 + \left( {\varepsilon - 1} \right)P_{n}^{m} \left( {\xi _{0}}
\right){\frac{{dP_{n}^{m} \left( {\xi _{0}}  \right)}}{{d\xi _{0}
}}}{\sum\limits_{q = m}^{\infty}  {S_{mnmq}^{\left( {2} \right)}
D_{mq}^{\left( {2} \right)}}} && = b_{mn}^{\left( {1} \right)}
\left( {\varepsilon - 1} \right)P_{n}^{m} \left( {\xi _{0}}
\right){\frac{{dP_{n}^{m} \left( {\xi _{0}}  \right)}}{{d\xi _{0}}} }, \nonumber \\
 \left( {\varepsilon {\frac{{dP_{n}^{m} \left( {\xi _{0}}  \right)}}{{d\xi
_{0}}} }Q_{n}^{m} \left( {\xi _{0}}  \right) - P_{n}^{m} \left( {\xi
_{0}} \right){\frac{{dQ_{n}^{m} \left( {\xi _{0}}  \right)}}{{d\xi
_{0}}} }}
\right)D_{mn}^{\left( {2} \right)} && \nonumber \\
 + \left( {\varepsilon - 1} \right)P_{n}^{m} \left( {\xi _{0}}
\right){\frac{{dP_{n}^{m} \left( {\xi _{0}}  \right)}}{{d\xi _{0}
}}}{\sum\limits_{q = m}^{\infty}  {S_{mnmq}^{\left( {1} \right)}
D_{mq}^{\left( {1} \right)}}} && = b_{mn}^{\left( {2} \right)}
\left( {\varepsilon - 1} \right)P_{n}^{m} \left( {\xi _{0}}
\right){\frac{{dP_{n}^{m} \left( {\xi _{0}}  \right)}}{{d\xi _{0}}}
}, \nonumber \\ \label{eq27}
 \end{eqnarray}

\noindent where the coefficients

\begin{eqnarray}
 a_{mn}^{\left( {j} \right)} &=& - \delta _{n1} f\left( {\delta _{m1} E_{0x} -
\delta _{m0} E_{0z}}  \right) + \left( { - 1} \right)^{j}\delta
_{n0} \delta
_{m0} E_{0z} {\frac{{l}}{{2}}}, \nonumber\\
 b_{mn}^{\left( {1} \right)} &=& b_{mn}^{\left( {2} \right)} = - \delta _{m1}
\delta _{n1} fE_{0y} , \label{eq28}
 \end{eqnarray}

\noindent define excitation field. In Eq.~(\ref{eq28}), $\delta
_{n1}$ is Kronecker delta symbol. It should be noted here that due
to axial symmetry of the considered cluster, the systems of
equations (\ref{eq26}) and (\ref{eq27}) allow one to find the
coefficients $C_{mn}^{\left( {j} \right)} $ and $D_{mn}^{\left( {j}
\right)}$ for given order $m$, while degree $n$ runs over $n=m$, $m
+ 1$, $m + 2$, \ldots , $m + N$, where $N$ is a large number that
defines accuracy of the solution.

Induced dipole moment of a cluster of two prolate nanospheroids,
placed in the field of a plane electromagnetic wave, can be
calculated by analogy with a single prolate nanospheroid
\cite{ref31}, that is, by finding far field asymptotes of the
potential (\ref{eq23}). As a result, for dipole moment induced in
the $j$-th nanospheroid ($j = 1$, 2) we have

\begin{equation}
d_{x}^{\left( {j} \right)} = {\frac{{2f_{}^{2}}}
{{3}}}C_{11}^{\left( {j} \right)} ,\quad d_{y}^{\left( {j} \right)}
= {\frac{{2f_{}^{2} }}{{3}}}D_{11}^{\left( {j} \right)} ,\quad
d_{z}^{\left( {j} \right)} = {\frac{{f_{}^{2}}} {{3}}}C_{01}^{\left(
{j} \right)} , \label{eq29}
\end{equation}

\noindent and the total dipole moment of the cluster will be the sum
of the momenta (\ref{eq29}). Scattering and absorption
cross-sections can be easily found if the dipole momenta
(\ref{eq29}) are known \cite{ref58}:

\begin{equation}
\sigma ^{abs} = 4\pi \left( {{\frac{{\omega}} {{v_{c}}} }}
\right){\frac{{ \textrm{Im} \left( {{\mathbf {d}}{\mathbf {E}}_{0}
^{
*}} \right)}}{{{\left| {{\mathbf {E}}_{0}}  \right|}^{2}}}},\quad
\sigma ^{scat} = {\frac{{8\pi}} {{3}}}\left( {{\frac{{\omega}}
{{v_{c}}} }} \right)^{4}{\frac{{{\left| {{\mathbf {d}}}
\right|}^{2}}}{{{\left| {{\mathbf {E}}_{0}}  \right|}^{2}}}},
\label{eq30}
\end{equation}

\noindent where ${\mathbf {d}} = {\mathbf {d}}^{\left( {1} \right)}
+ {\mathbf {d}}^{\left( {2} \right)}$ denotes dipole momentum of the
whole system, and asterix denotes an operation of the complex
conjugation.

\begin{figure}[there]
\includegraphics[width=8.0cm]{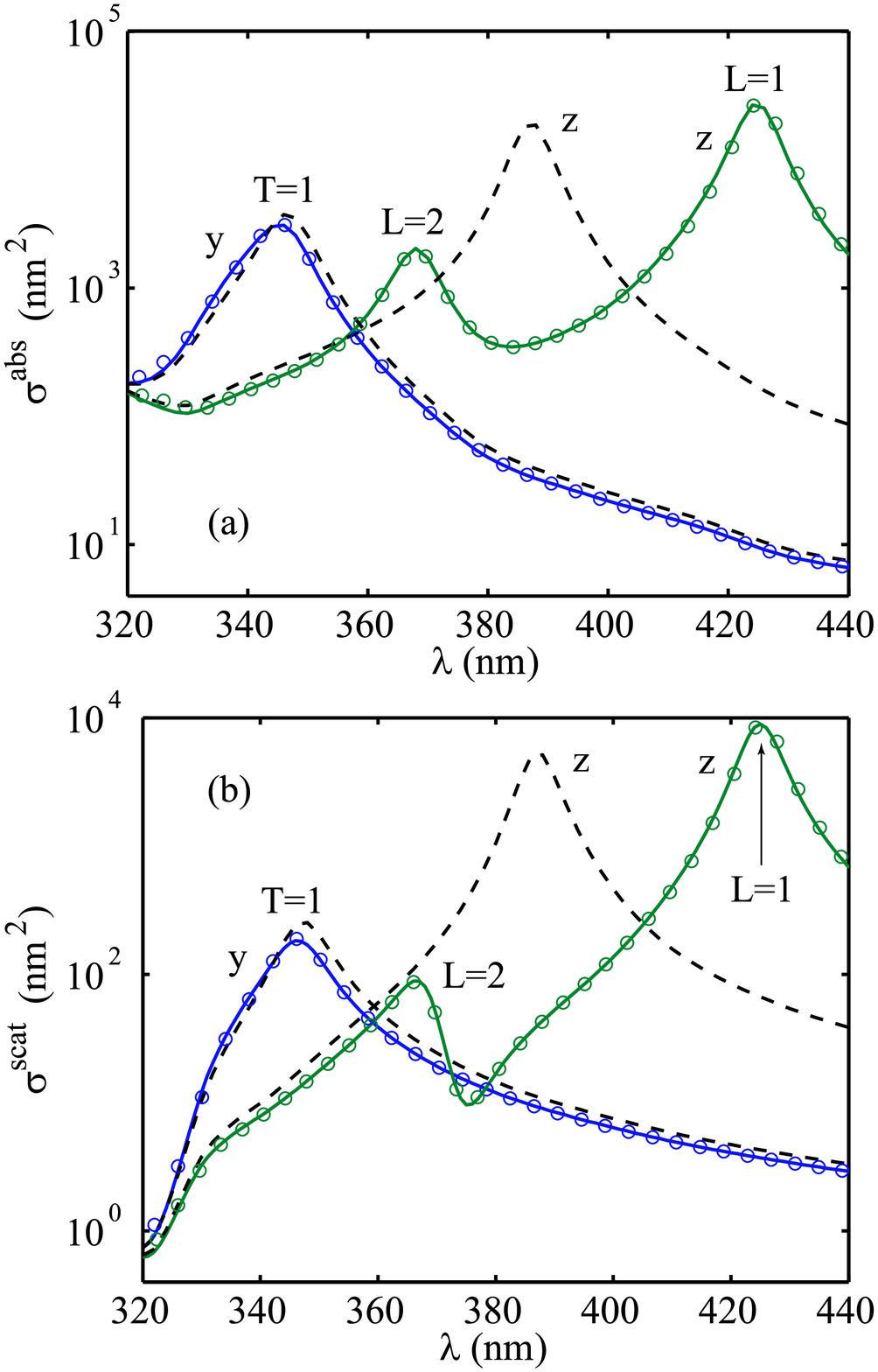}
\caption{\label{fig6} (Color online) Absorption (a) and scattering
(b) cross-sections of a cluster of two identical prolate
nanospheroids made from silver as a function of the wavelength. The
large semi-axes of the nanospheroids are $c=15$ nm, the aspect
ratios are $a/c=0.6$, and the distance between the nanospheroids'
centers is $l/2c = 1.05$. The labels x, y, z correspond to
polarization of an incident wave along the x, y, z axis. The circles
correspond to finite element simulations with Comsol
Multiphysics{\textregistered} software. Doubled absorption
cross-section and quadrupled scattering cross-section of a single
nanospheroid are shown by dashed curves.}
\end{figure}

In Fig.~\ref{fig6}, absorption (a) and scattering (b) cross-sections
of a cluster of two identical prolate nanospheroids made from
silver, are shown as a function of the wavelength. For longitudinal
(z) polarization, both of the cross-sections have two peaks that
correspond to longitudinal plasmonic oscillations with L = 1, 2 (see
Fig.~\ref{fig2}(b)). It is very important that both of the peaks are
split substantially relatively the case of a single spheroid (the
``z'' dashed curve) due to strong interaction between the
nanospheroids.

On the contrary, for transversal (x or y) polarization one can see
only one peak due to excitation of the symmetrical T = 1 mode, and
this peak is shifted just slightly relatively the single spheroid
resonance (the ``y'' dashed curve). It means that transversal (x or
y polarization) excitation of a two-spheroid cluster induces only
weak interaction between the nanospheroids (see the dispersion
curves for T-modes on Fig.~\ref{fig2}(a)). Due to this weak
interaction, absorption and scattering cross-sections are
approximately equal to doubled and quadrupled cross-section of a
single spheroid respectively.

It should be noticed that in Fig.~\ref{fig6} the maxima of
absorption, corresponding to plasmon oscillations of M-type that
should lie in the interval $\omega _{pl} / \sqrt {2} < \omega <
\omega _{pl} $, that corresponds to $326 < \lambda < 337$ nm for
silver \cite{ref49}, are not visible. It is related to the fact that
M-modes interact with a homogeneous electric field weakly and can be
effectively excited only by a source of radiation that is nonuniform
in comparison with a size of the gap between the nanoparticles
\cite{ref57} (see Fig.~\ref{fig3}).

To control correctness and accuracy of our analytical calculations,
we have also carried out finite element simulation of this system
with Comsol Multiphysics{\textregistered} software. The results of
this simulation are shown by the circles in Fig.~\ref{fig6}. One can
see that there is a fine agreement between the analytical and pure
numerical calculations. This fact confirms correctness and accuracy
of the both of approaches.

\subsection{\label{sect3b}A cluster of two oblate nanospheroids}

The case of two oblate nanospheroids is in many aspects similar to
the case of two prolate nanospheroids, considered above. So let us
again choose local systems of coordinates that are connected with
each of the nanospheroids and have origins in their centers (see
Fig.~\ref{fig1}(b)). The potential inside the $j$-th nanospheroid
again can be presented as a series in spheroidal harmonics
(\ref{eq16}), while the potential outside the oblate spheroids can
be presented in the form:

\begin{equation}
\varphi ^{out} = \varphi _{1}^{out} + \varphi _{2}^{out} + \varphi
_{0} , \label{eq31}
\end{equation}

\noindent where $\varphi _{1}^{out} ,\varphi _{2}^{out} $ are
contributions from the first and second nanospheroids (see
Eq.~(\ref{eq17})) and $\varphi _{0}$ is the potential of the
external electric field (\ref{eq22}). In the local coordinates of
the $j$-th ($j = 1$, 2) oblate nanospheroids, it looks like

\begin{eqnarray}
 \varphi _{0}^{\left( {j} \right)} = &&- if\left( {E_{0x} \cos \left( {\phi
_{j}}  \right) + E_{0y} \sin \left( {\phi _{j}}  \right)}
\right)P_{1}^{1}
\left( {i\xi _{j}}  \right)P_{1}^{1} \left( {\eta _{j}}  \right) \nonumber \\
 &&+ ifE_{0z} P_{1} \left( {i\xi _{j}}  \right)P_{1} \left( {\eta _{j}}
\right) + \left( { - 1} \right)^{j}E_{0x} {\frac{{l}}{{2}}} .
\label{eq32}
 \end{eqnarray}

Making use of the boundary conditions (\ref{eq25}) with $\xi _{0} =
c / \sqrt {a^{2} - c^{2}}  = c / f $ and the addition-translation
theorem (\ref{eq18}), we shall obtain the following systems of
equations ($n = 0$, 1, 2, \ldots ; $m = 0$, 1, 2, \ldots , $n$):

\begin{eqnarray}
 \left( {\varepsilon {\frac{{dP_{n}^{m} \left( {i\xi _{0}}  \right)}}{{d\xi
_{0}}} }Q_{n}^{m} \left( {i\xi _{0}}  \right) - P_{n}^{m} \left(
{i\xi _{0} } \right){\frac{{dQ_{n}^{m} \left( {i\xi _{0}}
\right)}}{{d\xi _{0}}} }}
\right)C_{mn}^{\left( {1} \right)}&& \nonumber \\
 + \left( {\varepsilon - 1} \right)P_{n}^{m} \left( {i\xi _{0}}
\right){\frac{{dP_{n}^{m} \left( {i\xi _{0}}  \right)}}{{d\xi _{0}
}}}{\sum\limits_{q = 0}^{\infty}  {{\sum\limits_{p = 0}^{q}
{M_{mnpq}^{\left( {2} \right)} C_{pq}^{\left( {2} \right)}}} } }&& =
a_{mn}^{\left( {1} \right)} \left( {\varepsilon - 1}
\right)P_{n}^{m} \left( {i\xi _{0}}  \right){\frac{{dP_{n}^{m}
\left( {i\xi _{0}}  \right)}}{{d\xi
_{0}}} }, \nonumber \\
 \left( {\varepsilon {\frac{{dP_{n}^{m} \left( {i\xi _{0}}  \right)}}{{d\xi
_{0}}} }Q_{n}^{m} \left( {i\xi _{0}}  \right) - P_{n}^{m} \left(
{i\xi _{0} } \right){\frac{{dQ_{n}^{m} \left( {i\xi _{0}}
\right)}}{{d\xi _{0}}} }}
\right)C_{mn}^{\left( {2} \right)}&& \nonumber \\
 + \left( {\varepsilon - 1} \right)P_{n}^{m} \left( {i\xi _{0}}
\right){\frac{{dP_{n}^{m} \left( {i\xi _{0}}  \right)}}{{d\xi _{0}
}}}{\sum\limits_{q = 0}^{\infty}  {{\sum\limits_{p = 0}^{q}
{M_{mnpq}^{\left( {1} \right)} C_{pq}^{\left( {1} \right)}}} } } &&=
a_{mn}^{\left( {2} \right)} \left( {\varepsilon - 1}
\right)P_{n}^{m} \left( {i\xi _{0}}  \right){\frac{{dP_{n}^{m}
\left( {i\xi _{0}}  \right)}}{{d\xi _{0}}} }, \nonumber \\
\label{eq33}
 \end{eqnarray}

\noindent and ($n = 1$, 2, 3, \ldots ; $m = 1$, 2, 3, \ldots , $n$)

\begin{eqnarray}
 \left( {\varepsilon {\frac{{dP_{n}^{m} \left( {i\xi _{0}}  \right)}}{{d\xi
_{0}}} }Q_{n}^{m} \left( {i\xi _{0}}  \right) - P_{n}^{m} \left(
{i\xi _{0} } \right){\frac{{dQ_{n}^{m} \left( {i\xi _{0}}
\right)}}{{d\xi _{0}}} }}
\right)D_{mn}^{\left( {1} \right)}&& \nonumber \\
 + \left( {\varepsilon - 1} \right)P_{n}^{m} \left( {i\xi _{0}}
\right){\frac{{dP_{n}^{m} \left( {i\xi _{0}}  \right)}}{{d\xi _{0}
}}}{\sum\limits_{q = 1}^{\infty}  {{\sum\limits_{p = 1}^{q}
{N_{mnpq}^{\left( {2} \right)} D_{pq}^{\left( {2} \right)}}} } } &&=
b_{mn}^{\left( {1} \right)} \left( {\varepsilon - 1}
\right)P_{n}^{m} \left( {i\xi _{0}}  \right){\frac{{dP_{n}^{m}
\left( {i\xi _{0}}  \right)}}{{d\xi
_{0}}} }, \nonumber \\
 \left( {\varepsilon {\frac{{dP_{n}^{m} \left( {i\xi _{0}}  \right)}}{{d\xi
_{0}}} }Q_{n}^{m} \left( {i\xi _{0}}  \right) - P_{n}^{m} \left(
{i\xi _{0} } \right){\frac{{dQ_{n}^{m} \left( {i\xi _{0}}
\right)}}{{d\xi _{0}}} }}
\right)D_{mn}^{\left( {2} \right)}&& \nonumber \\
 + \left( {\varepsilon - 1} \right)P_{n}^{m} \left( {i\xi _{0}}
\right){\frac{{dP_{n}^{m} \left( {i\xi _{0}}  \right)}}{{d\xi _{0}
}}}{\sum\limits_{q = 1}^{\infty}  {{\sum\limits_{p = 1}^{q}
{N_{mnpq}^{\left( {1} \right)} D_{pq}^{\left( {1} \right)}}} } } &&=
b_{mn}^{\left( {2} \right)} \left( {\varepsilon - 1}
\right)P_{n}^{m} \left( {i\xi _{0}}  \right){\frac{{dP_{n}^{m}
\left( {i\xi _{0}}  \right)}}{{d\xi _{0}}} }, \nonumber
\\ \label{eq34}
 \end{eqnarray}

\noindent where

\begin{eqnarray}
 a_{mn}^{\left( {j} \right)} &=& i\delta _{n1} f \left( {\delta _{m1}
E_{0x} - \delta _{m0} E_{0z}}  \right) + \left( { - 1} \right)^{j +
1}\delta
_{m0} \delta _{n0} E_{0x} {\frac{{l}}{{2}}}, \nonumber \\
 b_{mn}^{\left( {1} \right)} &=& b_{mn}^{\left( {2} \right)} = i\delta _{m1}
\delta _{n1} f E_{0y} . \label{eq35}
 \end{eqnarray}

Apparently, the equations (\ref{eq33}) and (\ref{eq34}) have more
complicated structure than Eqs.~(\ref{eq26}) and (\ref{eq27})
because now, due to lack of axial symmetry, one can not split the
system of equations into systems with fixed order $m$ of the
Legendre function.

For calculation of absorption and scattering cross-sections of a
cluster in the field of a plane electromagnetic wave, one can again
use Eq.~(\ref{eq30}), where dipole momenta of each spheroid can be
expressed through the solutions of Eqs.~(\ref{eq33}) and
(\ref{eq34}) ($j = 1$, 2)

\begin{equation}
d_{x}^{\left( {j} \right)} = - {\frac{{2f_{}^{2}}}
{{3}}}C_{11}^{\left( {j} \right)} ,\quad d_{y}^{\left( {j} \right)}
= - {\frac{{2f_{}^{2} }}{{3}}}D_{11}^{\left( {j} \right)} ,\quad
d_{z}^{\left( {j} \right)} = - {\frac{{f_{}^{2}}}
{{3}}}C_{01}^{\left( {j} \right)} . \label{eq36}
\end{equation}

\begin{figure}[there]
\includegraphics[width=8.0cm]{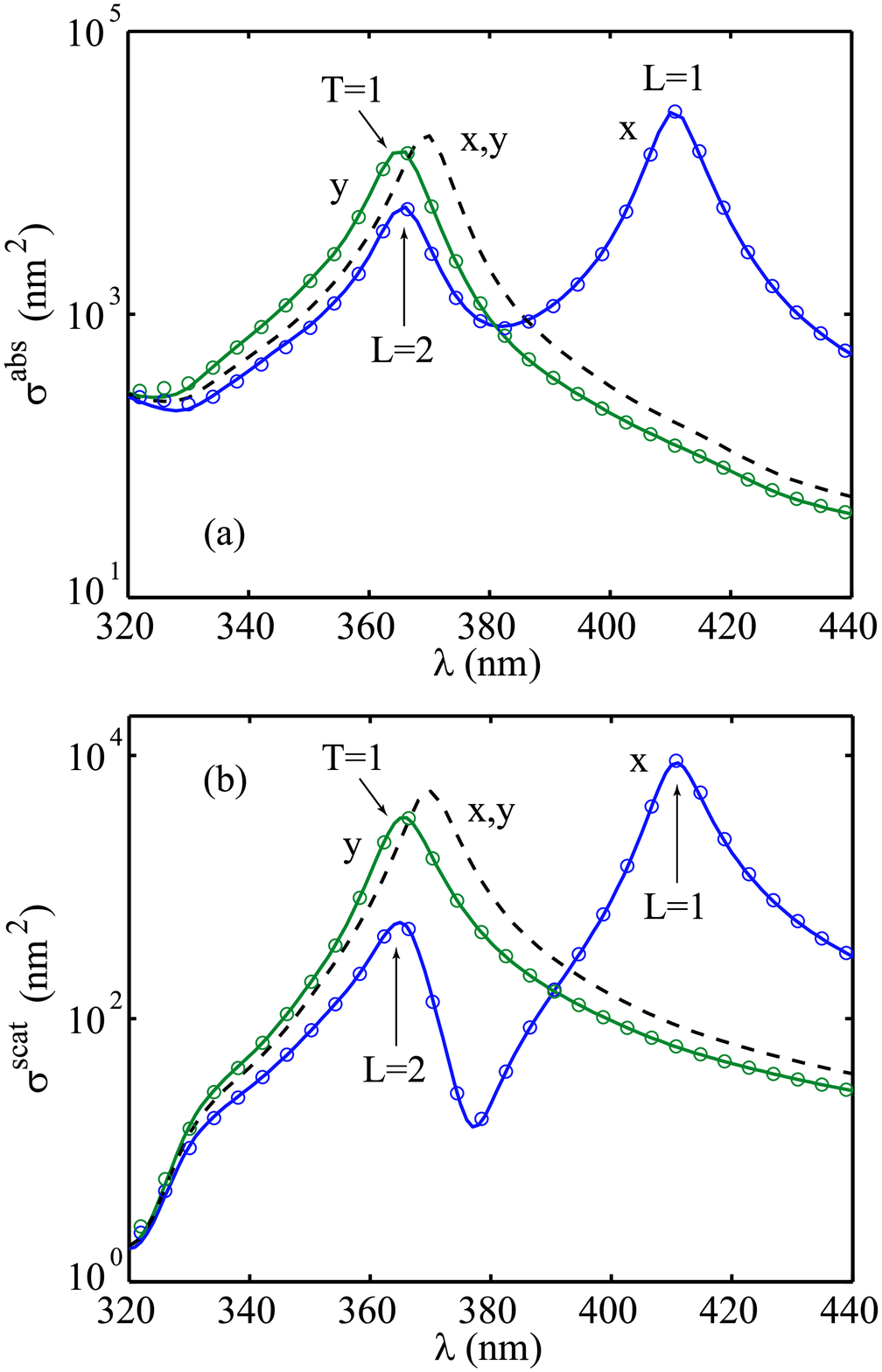}
\caption{\label{fig7} (Color online) Absorption (a) and scattering
(b) cross-sections of a cluster of two identical oblate
nanospheroids made from silver as a function of the wavelength. The
large semi-axes of the nanospheroids are $a=15$ nm, the aspect
ratios are $c/a=0.6$, the distance between the nanospheroids'
centers is $l/2a = 1.05$. The labels x, y, z correspond to
polarization of an incident wave along the x, y, z axis. The circles
correspond to finite element simulations with Comsol
Multiphysics{\textregistered} software. Doubled absorption
cross-section and quadrupled scattering cross-section of a single
nanospheroid are shown by dashed curves.}
\end{figure}

In Fig.~\ref{fig7}, the absorption (a) and scattering (b)
cross-sections of a cluster of two identical oblate nanospheroids
made from silver are shown as a function of wavelength. For
longitudinal (x) polarization, both the cross-sections have two
peaks, that correspond to antisymmetric plasmonic oscillations with
L = 1, 2 (see Fig.~\ref{fig4}(b)). It is very important that now
only one peak (L = 1) is shifted substantially relatively the case
of a single spheroid (the ``x, y'' dashed curves) due to strong
interaction between the nanospheroids. The L = 2 mode suffers only a
small shift in agreement with Fig.~\ref{fig4}(b).

For transversal (y) polarization, one can see only one peak due to
excitation of the symmetrical T =1 mode, and this peak is only
slightly shifted relatively the single spheroid resonance (the
dashed curve). It means that transversal (y polarization) excitation
of a two-spheroid cluster results only in weak interaction between
the nanospheroids (see the dispersion curves for the T-modes on
Fig.~\ref{fig4}(a)). Due to this weak interaction, absorption and
scattering cross-sections for this polarization are approximately
equal to doubled and quadrupled cross-sections of a single spheroid
respectively. It is also interesting that plasmonic frequency of the
L = 2 mode is very close to plasmonic frequency of the T = 1 mode.
This fact can be easily understood from analysis of Fig.~\ref{fig4}.
Indeed, when width of the gap tends to zero, plasmonic frequency of
L = 2 modes also decreases to zero, while plasmonic frequency of the
T = 1 mode increases slightly. So, at some point these modes will
intersect and have the same frequencies, and we observe this
situation in Fig.~\ref{fig7}.

It should be noticed that in Fig.~\ref{fig7} the maxima of
absorption corresponding to plasmon oscillations of M-type again are
not visible. It is related to the fact that M-modes interact with a
homogeneous electric field weakly and can be effectively excited
only by a source of radiation that is nonuniform in comparison with
a size of the gap between nanoparticles \cite{ref57} (see
Fig.~\ref{fig3}).

To control correctness and accuracy of our analytical calculations
for a cluster of two oblate spheroids, we have also carried out
finite element simulation of this system with Comsol
Multiphysics{\textregistered} software. The results of this
simulation are shown by the circles in Fig.~\ref{fig7}. One can see
that there is a fine agreement between the analytical and pure
numerical calculations. This fact confirms correctness and accuracy
of the both of approaches again.

\subsection{\label{sect3c}Enhancement of local fields}

The most important characteristic of nanoparticles clusters is the
factor of incident field enhancement in the gap between the
nanoparticles. It is the characteristic that allows determining the
excitation rate of molecules near the nanoparticles or intensity of
surface-enhanced Raman scattering \cite{ref4}. Moreover, achievement
of high values of this factor is the main goal of optical
nano-antennas development.

\begin{figure}[there]
\includegraphics[width=8.0cm]{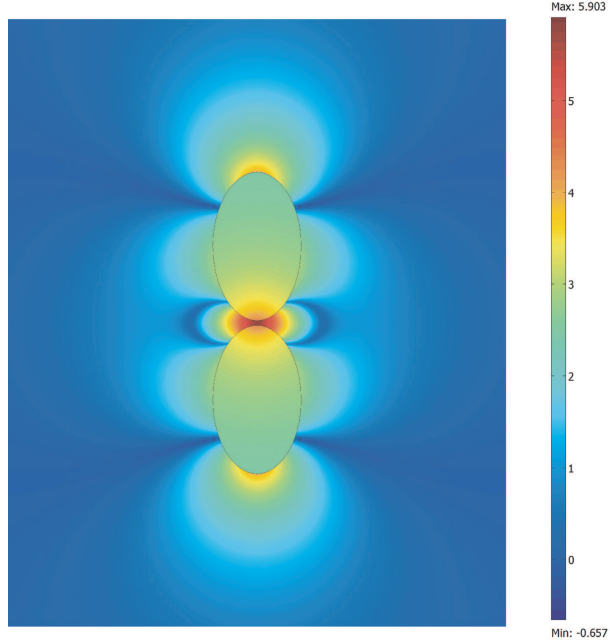}
\caption{\label{fig8} (Color online) Spatial distribution of
${\left| {{\mathbf {E}}} \right|}^{2} / {\left| {{\mathbf {E}}_{0}}
\right|}^{2}$ in a cluster of two prolate nanospheroids for $\lambda
= 425$ nm at longitudinal polarization of the excitation field (L =
1 plasmonic resonance, common logarithmic scale).}
\end{figure}

Distribution of squared electric field for the L=1 resonance in a
cluster of two prolate spheroids is shown in Fig.~\ref{fig8}, that
shows that, indeed, the maximal field enhancement takes place in the
gap between the nanoparticles on their surfaces. The field maxima
are also present on the outer side of the cluster; however, field
amplitude is essentially less there. According to general theorems
for harmonic functions, the field maximum can be reached only on the
region boundaries. In our case, the field maxima are reached in
those points of the spheroids' surface where the distance between
the spheroids is minimal.

Using Eqs.~(\ref{eq10}) and (\ref{eq17}), one can find explicit
expressions for the field enhancement factor $G$. For clusters of
two identitcal prolate spheroids in the considered configuration
(Fig.~\ref{fig1}(a)), one can obtain the following expression for
the field maximum in the case of an incident field polarized along
the z-axis:

\begin{equation}
G = {\frac{{{\left| {{\mathbf {E}}} \right|}^{2}}}{{{\left| {E_{0z}}
\right|}^{2}}}} = {\left| {1 - {\frac{{1}}{{E_{0z}
f}}}{\sum\limits_{n = 1}^{\infty}  {C_{0n} \left( {{\frac{{dQ_{n}
\left( {\xi _{1}} \right)}}{{d\xi _{1}}} } + {\frac{{dQ_{n} \left(
{\xi _{2}}  \right)}}{{d\xi _{2}}} }} \right)}}}  \right|}^{2},
\label{eq37}
\end{equation}

\noindent where $\xi _{1} = c / f$ and $\xi _{2} = \left( {l - c}
\right) / f$, while $C_{0n} = C_{0n}^{\left( {1} \right)} = - \left(
{ - 1} \right)^{n}C_{0n}^{\left( {2} \right)} $. In the most
interesting case of a small gap and strongly prolate spheroids, $\xi
_{1}$, $\xi _{2} \approx 1$, and one may use the asymptotic form
${\left. {{\frac{{dQ_{n} \left( {\xi} \right)}}{{d\xi}} }}
\right|}_{\xi \approx 1} \approx - {\frac{{1}}{{2\left( {\xi - 1}
\right)}}}$. As a result, the field enhancement factor takes the
form:

\begin{equation}
G = {\frac{{{\left| {{\mathbf {E}}} \right|}^{2}}}{{{\left| {E_{0z}}
\right|}^{2}}}} \approx {\left| {{\frac{{1}}{{2E_{0z} f}}}\left(
{{\frac{{1}}{{\xi _{1} - 1}}} + {\frac{{1}}{{\xi _{2} - 1}}}}
\right){\sum\limits_{n = 1}^{\infty} {C_{0n}}} }  \right|}^{2}.
\label{eq38}
\end{equation}

In the case of clusters of two identical oblate spheroids
(Fig.~\ref{fig1}(b)) and incident field polarized along the x-axis,
we obtain

\begin{eqnarray}
 G = {\frac{{{\left| {{\mathbf {E}}} \right|}^{2}}}{{{\left| {E_{0x}}
\right|}^{2}}}} = {\left| {{\frac{{1}}{{E_{0x} f}}}{\sum\limits_{n =
1}^{\infty}  {{\sum\limits_{m = 0}^{n} {C_{mn} {\left.
{{\frac{{dP_{n}^{m} \left( {\eta}  \right)}}{{d\eta}} }}
\right|}_{\eta = 0} \left( {{\frac{{1}}{{\xi _{1}}} }Q_{n}^{m}
\left( {i\xi _{1}}  \right) - {\frac{{1}}{{\xi _{2}}} }Q_{n}^{m}
\left( {i\xi _{2}}  \right)} \right)}}}
}} \right|}^{2} \nonumber \\
 + {\left| {1 - {\frac{{1}}{{E_{0x} f}}}{\sum\limits_{n = 1}^{\infty}
{{\sum\limits_{m = 0}^{n} {C_{mn} P_{n}^{m} \left( {0} \right)\left(
{\sqrt {1 + {\frac{{1}}{{\xi _{1}^{2}}} }} {\frac{{dQ_{n}^{m} \left(
{i\xi _{1}} \right)}}{{d\xi _{1}}} } + \sqrt {1 + {\frac{{1}}{{\xi
_{2}^{2}}} }} {\frac{{dQ_{n}^{m} \left( {i\xi _{2}}  \right)}}{{d\xi
_{2}}} }} \right)}}} }} \right|}^{2}, \label{eq39}
 \end{eqnarray}

\noindent where $\xi _{1} = \sqrt {a^{2} / f^{2} - 1} $ è $\xi _{2}
= \sqrt {\left( {l - a} \right)^{2} / f^{2} - 1} $; $C_{mn} = -
\left( { - 1} \right)^{m}C_{mn}^{\left( {1} \right)} =
C_{mn}^{\left( {2} \right)}$.

\begin{figure}[there]
\includegraphics[width=8.0cm]{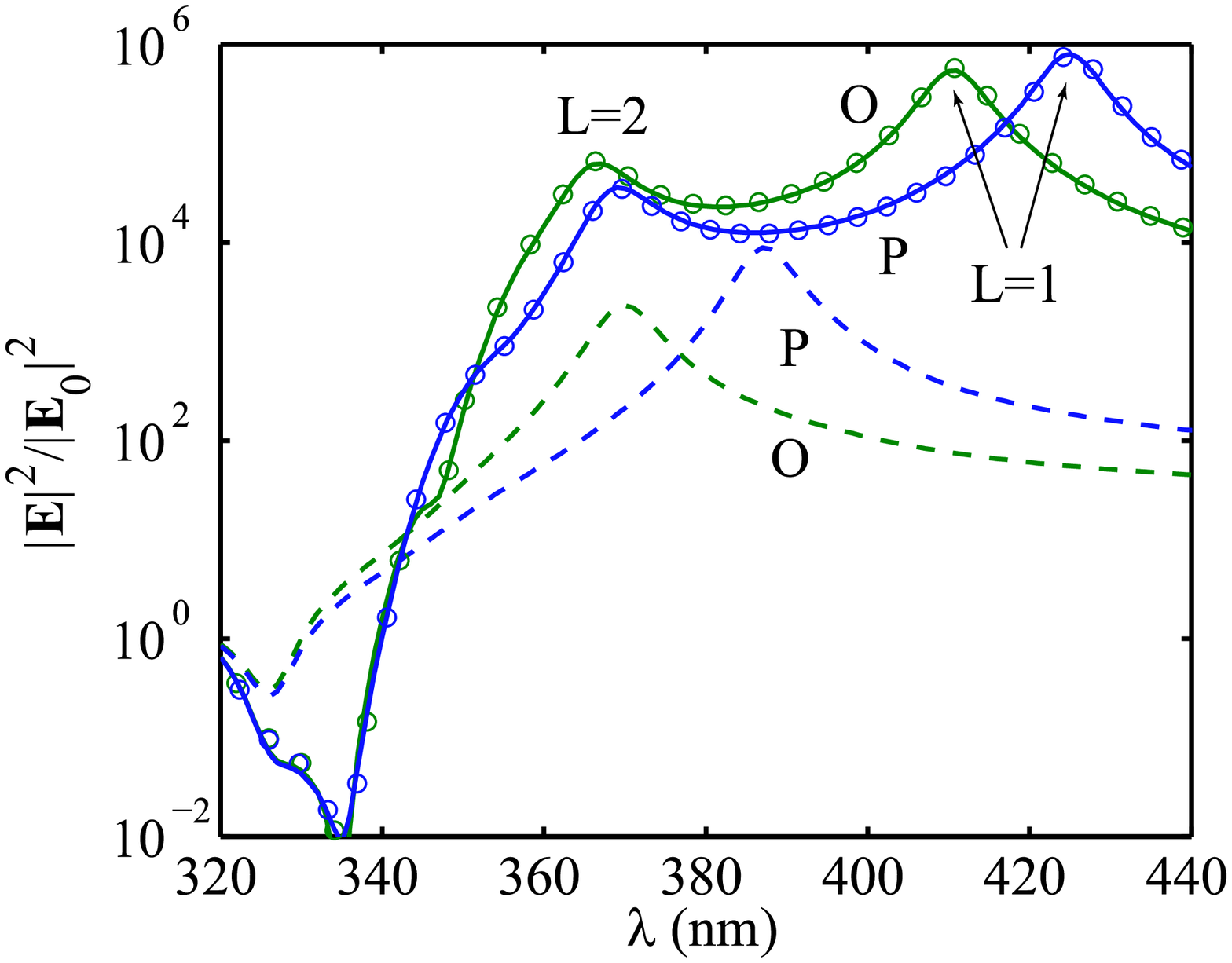}
\caption{\label{fig9} (Color online) Enhancement of ${\left|
{{\mathbf {E}}} \right|}^{2}$ in the gap between two identical
silver nanospherodis as a function of the wavelength. P and O labels
correspond to prolate and oblate spheroids correspondigly. The large
semi-axes of the nanospheroids are 15 nm, the aspect ratios are 0.6,
$l/2c = 1.05$ for a cluster of prolate spheroids and $l/2a = 1.05$
for a cluster of oblate spheroids. Enhancements for single
nanospheroids are shown by the dashed curve.}
\end{figure}

In Fig.~\ref{fig9}, the dependence of squared electric field
enhancement (\ref{eq37}) and (\ref{eq39}) for clusters of two
identical silver nanospheroids on the wavelength, is shown.
Comparing peaks positions to the dispersion curves in
Fig.~\ref{fig2} and Fig.~\ref{fig4}, one can come to a conclusion
that only ``L'' types of plasmon modes are excited in the clusters
for the considered configurations of nanospheroids and incident
electromagnetic wave polarizations (along the line joining the
nanoparticles' centers). In particular, excitation of the L = 1 and
L = 2 modes is noticeable. At that, the position of squared field
enhancement peaks agrees with the maxima of the absorption and
scattering cross-sections shown in Fig.~\ref{fig6} and \ref{fig7} by
the solid lines z and x correspondingly. It should be mentioned that
the value of squared field enhancement near a cluster of two
nanospheroids can reach up to $10^{6}$. In the case of single
nanoparticles, this value is almost two orders less than that of
clusters (cf. the solid and dashed curves in Fig.~\ref{fig9}). This
fact determines a greater attractiveness of metal nanoparticles
clusters in comparison to single nanoparticles for investigation of
Surface Enhance Raman Scattering (SERS) and Surface-Enhanced
Fluorescence (SEF). Let us note that the obtained great field
enhancement factors can be slightly less in practice since for small
particles and for small gaps between them, non-local and other
effects not considered in this research become essential.

\section{\label{sect4}A cluster of two nanospheroids in the field of a dipole
source}

In the previous section, we have considered the case of a
nano-antenna placed in the field of a plane wave. However, highly
nonuniform optical fields occur very often in nano-environment. For
example, such fields arise when a plasmonic nano-antenna is excited
by an atom, or a molecule, or another nanolocalized source of light.
So, in this section, we will consider an important case of a
two-nanospheroid cluster in the field of electric dipole sources.
Excitation of the cluster by magnetic dipole and electric quadrupole
sources can be analyzed analogously.

\subsection{\label{sect4a}A cluster of two prolate nanospheroids}

A case of two prolate nanospheroids in the field of a dipole source
of radiation can be considered by perfect analogy with a case of the
same cluster in a uniform field. One should again look for solution
in the form (\ref{eq9}), (\ref{eq10}) and then apply the boundary
conditions (\ref{eq25}). The only difference is that now the
external potential is the potential $\varphi _{0}$ of the dipole
that has the following form in the $j$-th local system of
coordinates of a prolate spheroid \cite{ref55} ($j = 1$, 2):

\begin{eqnarray}
 &&\varphi _{0}^{\left( {j} \right)} = \left( {{\mathbf {d}}_{0} {\nabla
}'_{j}}  \right){\frac{{1}}{{{\left| {{\mathbf {r}} - {\mathbf
{{r}'}}}
\right|}}}} \nonumber \\
 &&= {\sum\limits_{n = 0}^{\infty}  {{\sum\limits_{m = 0}^{n} {{\left\{
{{\begin{array}{*{20}c}
 {P_{n}^{m} \left( {\xi _{j}}  \right)P_{n}^{m} \left( {\eta _{j}}
\right){\left[ {\left( {{\mathbf {d}}_{0} {\nabla} '_{j}}
\right)\alpha _{mn}^{\left( {j} \right)} \cos \left( {m\phi _{j}}
\right) + \left( {{\mathbf {d}}_{0} {\nabla} '_{j}}  \right)\beta
_{mn}^{\left( {j} \right)} \sin \left( {m\phi _{j}}  \right)}
\right]},} \hfill & {\xi _{j} < {\xi
}'_{j} ,} \hfill \\
 {Q_{n}^{m} \left( {\xi _{j}}  \right)P_{n}^{m} \left( {\eta _{j}}
\right){\left[ {\left( {{\mathbf {d}}_{0} {\nabla} '_{j}}
\right)\gamma _{mn}^{\left( {j} \right)} \cos \left( {m\phi _{j}}
\right) + \left( {{\mathbf {d}}_{0} {\nabla} '_{j}}  \right)\sigma
_{mn}^{\left( {j} \right)} \sin \left( {m\phi _{j}}  \right)}
\right]},} \hfill & {\xi _{j} > {\xi
}'_{j} .} \hfill \\
\end{array}}}  \right.}}}} } \nonumber \\  \label{eq40}
\end{eqnarray}

\noindent In Eq.~(\ref{eq40}), ${\mathbf {d}}_{0} $ denotes the
dipole momentum of a source placed at ${\mathbf {{r}'}}$, ${\nabla}
'_{j}$ is a gradient over ${\mathbf {{r}'}}$ in local coordinates,
and

\begin{eqnarray}
 {\left. {{\begin{array}{*{20}c}
 {\alpha _{mn}^{\left( {j} \right)}}  \hfill \\
 {\beta _{mn}^{\left( {j} \right)}}  \hfill \\
\end{array}}}  \right\}}  = {\frac{{1}}{{f_{}}} }\left( {2 - \delta _{m0}}
\right)\left( { - 1} \right)^{m}\left( {2n + 1} \right){\left[
{{\frac{{\left( {n - m} \right)!}}{{\left( {n + m} \right)!}}}}
\right]}^{2}Q_{n}^{m} \left( {{\xi} '_{j}}  \right)P_{n}^{m} \left(
{{\eta }'_{j}}  \right){\left\{ {{\begin{array}{*{20}c}
 {\cos \left( {m{\phi} '_{j}}  \right)} \hfill \\
 {\sin \left( {m{\phi} '_{j}}  \right)} \hfill \\
\end{array}}}  \right.}, \nonumber \\
 {\left. {{\begin{array}{*{20}c}
 {\gamma _{mn}^{\left( {j} \right)}}  \hfill \\
 {\sigma _{mn}^{\left( {j} \right)}}  \hfill \\
\end{array}}}  \right\}}  = {\frac{{1}}{{f_{}}} }\left( {2 - \delta _{m0}}
\right)\left( { - 1} \right)^{m}\left( {2n + 1} \right){\left[
{{\frac{{\left( {n - m} \right)!}}{{\left( {n + m} \right)!}}}}
\right]}^{2}P_{n}^{m} \left( {{\xi} '_{j}}  \right)P_{n}^{m} \left(
{{\eta }'_{j}}  \right){\left\{ {{\begin{array}{*{20}c}
 {\cos \left( {m{\phi} '_{j}}  \right)} \hfill \\
 {\sin \left( {m{\phi} '_{j}}  \right)} \hfill \\
\end{array}}}  \right.}, \label{eq41}
 \end{eqnarray}

\noindent are expansion coefficients of the unit charge potential in
local coordinates of a prolate spheroid.

As a result of applying of the boundary conditions, one can obtain a
system of equations for the unknown coefficients $C_{mn}^{\left( {j}
\right)} $, $D_{mn}^{\left( {j} \right)}$ in Eq.~(\ref{eq10}). The
new system can be easily derived from Eqs.~(\ref{eq26}) and
(\ref{eq27}) if one makes the following replacement for the
coefficients $a_{mn}^{\left( {j} \right)} $ and $b_{mn}^{\left( {j}
\right)}$:

\begin{eqnarray}
 &&a_{mn}^{\left( {1} \right)} = - \left( {{\mathbf {d}}_{0} {\nabla} '_{1}}
\right)\alpha _{mn}^{\left( {1} \right)} ,\quad a_{mn}^{\left( {2}
\right)} = - \left( {{\mathbf {d}}_{0} {\nabla} '_{2}} \right)\alpha
_{mn}^{\left(
{2} \right)} , \nonumber \\
 &&b_{mn}^{\left( {1} \right)} = - \left( {{\mathbf {d}}_{0} {\nabla} '_{1}}
\right)\beta _{mn}^{\left( {1} \right)} ,\quad b_{mn}^{\left( {2}
\right)} = - \left( {{\mathbf {d}}_{0} {\nabla} '_{2}} \right)\beta
_{mn}^{\left( {2} \right)} , \label{eq42}
 \end{eqnarray}

\noindent where $\alpha _{mn}^{\left( {j} \right)} $ and $\beta
_{mn}^{\left( {j} \right)}$ are defined by Eq.~(\ref{eq41}).

After Eqs.~(\ref{eq26}) and (\ref{eq27}) have been solved with
taking (\ref{eq42}) into account, one can find the total induced
dipole moment of the prolate nanospheroids

\begin{eqnarray}
 d_{x} &=& {\frac{{2f_{}^{2}}} {{3}}}\left( {C_{11}^{\left( {1} \right)} +
C_{11}^{\left( {2} \right)}}  \right), \nonumber \\
 d_{y} &=& {\frac{{2f_{}^{2}}} {{3}}}\left( {D_{11}^{\left( {1} \right)} +
D_{11}^{\left( {2} \right)}}  \right), \nonumber \\
 d_{z} &=& {\frac{{f_{}^{2}}} {{3}}}\left( {C_{01}^{\left( {1} \right)} +
C_{01}^{\left( {2} \right)}}  \right). \label{eq43}
 \end{eqnarray}

Knowing the dipole momenta (\ref{eq43}), it is easy to find
radiative decay rate of a dipole placed near the prolate
nanospheroid \cite{ref59}:

\begin{equation}
\gamma = {\frac{{P_{}^{rad}}} {{\hbar \omega}} } = {\frac{{\omega
^{3}}}{{3\hbar v_{c}^{3}}} }{\left| {{\mathbf {d}}_{0} + {\mathbf
{d}}} \right|}^{2}. \label{eq44}
\end{equation}

\noindent where $P^{rad}$ is radiation power at frequency $\omega$
and $\hbar \omega $ is emitted photon energy.

Radiative decay rate is a very important characteristics in such
applications as surface enhanced Raman scattering, surface-enhanced
fluorescence, nanolasers, and so on. To characterize the radiative
decay rate, it is naturally to normalize it to radiative decay rate
of a dipole in free space, $\gamma _{0} = {\frac{{P_{0}^{rad}}}
{{\hbar \omega}} } = {\frac{{\omega ^{3}}}{{3\hbar v_{c}^{3}}}
}{\left| {{\mathbf {d}}_{0}}  \right|}^{2}$.

\begin{figure}[there]
\includegraphics[width=8.0cm]{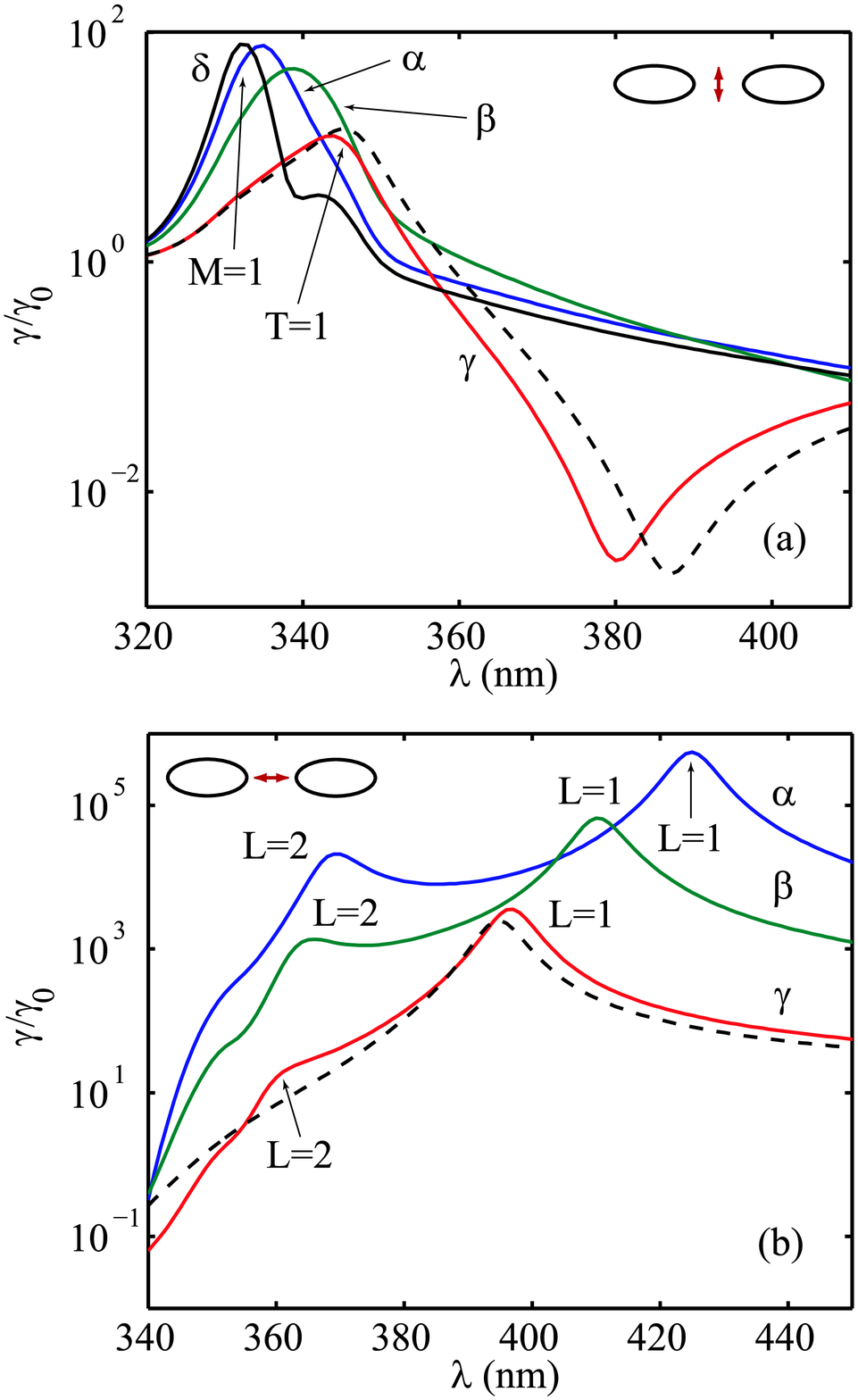}
\caption{\label{fig10} (Color online) Normalized radiative decay
rate of a dipole placed at the middle point between two identical
prolate nanospheroids made from silver as a function of the
wavelength. The dipole source moment is oriented along the x or y
axes (a) and along the z axis (b). The large semi-axes of the
nanospheroids are $c=15$ nm, the aspect ratios are $a/c=0.6$. The
curves $\alpha $, $\beta $, $\gamma $, $\delta $ correspond to $l/2c
= 1.05$, 1.1, 1.3 and 1.03, respectively. The asymptotic expression
obtained by approximation of the spheroids by point dipoles ($l/2c =
1.3$) is shown by the dashed curve.}
\end{figure}

In Fig.~\ref{fig10}, normalized radiative decay rate of a dipole
source placed at the middle point of the gap is shown. As it is well
seen in Fig.~\ref{fig10}(a), if the distance between the prolate
nanospheroids is small (Fig.~\ref{fig10}(a), the curves $\alpha$ and
$\delta $), the dipole source with a moment oriented perpendicular
to the cluster's axis of rotation, can excite both symmetrical T-
and M-modes. This fact differs from the case of excitation of the
same cluster with a plane wave, when M-modes with peaks located in
the region of $\lambda < 337$ nm (see Fig.~\ref{fig2}(a)) are not
excited. When the distance between the nanospheroids increases (see
Fig.~\ref{fig10}(a)), the peak corresponding to M-modes shifts to
$\lambda \approx 337$ nm ($\omega \approx \omega _{pl} / \sqrt {2}
$) and then disappear. After that point, only the peaks
corresponding to plasmonic T-modes can be observed. Of course, this
picture is in agreement with the behavior of the plasmonic M-modes
shown in Fig.~\ref{fig2}(a). We shall also notice that for large
enough distances between the nanospheroids (see Fig.~\ref{fig10}(a),
the curve $\gamma $) the self-consistent model \cite{ref3,ref21}, in
which nanoparticles are replaced by point dipoles with corresponding
polarizabilities \cite{ref3,ref34}, can be effectively used for
calculation of radiative decay rate (the dashed curve).

When dipole moment of a source is oriented along the axis of
symmetry (Fig.~\ref{fig10}(b)), only antisymmetric L-modes can be
excited due to symmetry reasons. From Fig.~\ref{fig10}(b), one can
also see that for small enough distances between the spheroids there
are two plasmonic modes (L = 1, 2) that interact with the dipole
source. When the distance between the spheroids diminishes, further
peaks of radiation power shift towards long wavelengths. At large
distances between the nanospheroids, there is only one maximum
corresponding to the L = 1 plasmonic mode (see Fig.~\ref{fig10}(b),
the curve $\gamma $). In this case, the radiative decay rate of a
dipole placed near a cluster of two prolate nanospheroids can be
also calculated with making use of the self-consistent analytical
model in which the spheroids are approximated by point dipoles (see
the dashed curve in Fig.~\ref{fig10}(b)).

\subsection{\label{sect4b}A cluster of two oblate nanospheroids}

A case of two oblate nanospheroids in the field of a dipole source
of radiation can be considered by perfect analogy with a case of the
same cluster in a uniform field. One should again look for solution
in the forms (\ref{eq16}) and (\ref{eq17}) and then apply the
boundary conditions (\ref{eq25}). The only difference is that now
the external potential is the potential $\varphi _{0}$ of the dipole
that has the following form in the $j$-th local system of
coordinates of an oblate spheroid \cite{ref55} ($j = 1$, 2):

\begin{eqnarray}
 &&\varphi _{0}^{\left( {j} \right)} = \left( {{\mathbf {d}}_{0} {\nabla
}'_{j}}  \right){\frac{{1}}{{{\left| {{\mathbf {r}} - {\mathbf
{{r}'}}}
\right|}}}} \nonumber \\
 &&= {\sum\limits_{n = 0}^{\infty}  {{\sum\limits_{m = 0}^{n} {{\left\{
{{\begin{array}{*{20}c}
 {P_{n}^{m} \left( {i\xi _{j}}  \right)P_{n}^{m} \left( {\eta _{j}}
\right){\left[ {\left( {{\mathbf {d}}_{0} {\nabla} '_{j}}
\right)\alpha _{mn}^{\left( {j} \right)} \cos \left( {m\phi _{j}}
\right) + \left( {{\mathbf {d}}_{0} {\nabla} '_{j}}  \right)\beta
_{mn}^{\left( {j} \right)} \sin \left( {m\phi _{j}}  \right)}
\right]},} \hfill & {\xi _{j} < {\xi
}'_{j} ,} \hfill \\
 {Q_{n}^{m} \left( {i\xi _{j}}  \right)P_{n}^{m} \left( {\eta _{j}}
\right){\left[ {\left( {{\mathbf {d}}_{0} {\nabla} '_{j}}
\right)\gamma _{mn}^{\left( {j} \right)} \cos \left( {m\phi _{j}}
\right) + \left( {{\mathbf {d}}_{0} {\nabla} '_{j}}  \right)\sigma
_{mn}^{\left( {j} \right)} \sin \left( {m\phi _{j}}  \right)}
\right]},} \hfill & {\xi _{j} > {\xi
}'_{j} .} \hfill \\
\end{array}}}  \right.}}}} } \nonumber \\  \label{eq45}
 \end{eqnarray}

\noindent In Eq.~(\ref{eq45}), ${\mathbf {d}}_{0}$ denotes the
dipole momentum of a source placed at ${\mathbf {{r}'}}$, ${\nabla}
'_{j}$ is gradient over ${\mathbf {{r}'}}$ in local coordinates, and

\begin{eqnarray}
 {\left. {{\begin{array}{*{20}c}
 {\alpha _{mn}^{\left( {j} \right)}}  \hfill \\
 {\beta _{mn}^{\left( {j} \right)}}  \hfill \\
\end{array}}}  \right\}}  = {\frac{{i}}{{f_{}}} }\left( {2 - \delta _{m0}}
\right)\left( { - 1} \right)^{m}\left( {2n + 1} \right){\left[
{{\frac{{\left( {n - m} \right)!}}{{\left( {n + m} \right)!}}}}
\right]}^{2}Q_{n}^{m} \left( {i{\xi} '_{j}}  \right)P_{n}^{m} \left(
{{\eta }'_{j}}  \right){\left\{ {{\begin{array}{*{20}c}
 {\cos \left( {m{\phi} '_{j}}  \right)} \hfill \\
 {\sin \left( {m{\phi} '_{j}}  \right)} \hfill \\
\end{array}}}  \right.}, \nonumber \\
 {\left. {{\begin{array}{*{20}c}
 {\gamma _{mn}^{\left( {j} \right)}}  \hfill \\
 {\sigma _{mn}^{\left( {j} \right)}}  \hfill \\
\end{array}}}  \right\}}  = {\frac{{i}}{{f_{}}} }\left( {2 - \delta _{m0}}
\right)\left( { - 1} \right)^{m}\left( {2n + 1} \right){\left[
{{\frac{{\left( {n - m} \right)!}}{{\left( {n + m} \right)!}}}}
\right]}^{2}P_{n}^{m} \left( {i{\xi} '_{j}}  \right)P_{n}^{m} \left(
{{\eta }'_{j}}  \right){\left\{ {{\begin{array}{*{20}c}
 {\cos \left( {m{\phi} '_{j}}  \right)} \hfill \\
 {\sin \left( {m{\phi} '_{j}}  \right)} \hfill \\
\end{array}}}  \right.}, \label{eq46}
 \end{eqnarray}

\noindent are expansion coefficients of the unit charge potential in
local coordinates of an oblate spheroid.

As a result of applying of the boundary conditions, one can obtain
the system of equations for the unknown coefficients $C_{mn}^{\left(
{j} \right)} $, $D_{mn}^{\left( {j} \right)}$ in Eq.~(\ref{eq17}).
The system of equations for these coefficients can be easily derived
from Eqs.~(\ref{eq33}) and (\ref{eq34}) if one makes the following
replacement for the coefficients $a_{mn}^{\left( {j} \right)} $ and
$b_{mn}^{\left( {j} \right)} $

\begin{eqnarray}
 &&a_{mn}^{\left( {1} \right)} = - i\left( {{\mathbf {d}}_{0} {\nabla} '_{1}}
\right)\alpha _{mn}^{\left( {1} \right)} ,\quad a_{mn}^{\left( {2}
\right)} = - i\left( {{\mathbf {d}}_{0} {\nabla} '_{21}}
\right)\alpha _{mn}^{\left(
{2} \right)} ,\nonumber \\
 &&b_{mn}^{\left( {1} \right)} = - i\left( {{\mathbf {d}}_{0} {\nabla} '_{1}}
\right)\beta _{mn}^{\left( {1} \right)} ,\quad b_{mn}^{\left( {2}
\right)} = - i\left( {{\mathbf {d}}_{0} {\nabla} '_{2}} \right)\beta
_{mn}^{\left( {2} \right)} , \label{eq47}
 \end{eqnarray}

\noindent where $\alpha _{mn}^{\left( {j} \right)} $ and $\beta
_{mn}^{\left( {j} \right)} $ are defined in Eq.~(\ref{eq46}).

The total induced dipole moment of a cluster of two oblate
nanospheroids can be found by using the following expressions (cf.
Eq.~(\ref{eq43})):

\begin{eqnarray}
 d_{x} &=& - {\frac{{2f_{}^{2}}} {{3}}}\left( {C_{11}^{\left( {1} \right)} +
C_{11}^{\left( {2} \right)}}  \right), \nonumber \\
 d_{y} &=& - {\frac{{2f_{}^{2}}} {{3}}}\left( {D_{11}^{\left( {1} \right)} +
D_{11}^{\left( {2} \right)}}  \right), \nonumber \\
 d_{z} &=& - {\frac{{f_{}^{2}}} {{3}}}\left( {C_{01}^{\left( {1} \right)} +
C_{01}^{\left( {2} \right)}}  \right). \label{eq48}
 \end{eqnarray}

\begin{figure}[there]
\includegraphics[width=7.5cm]{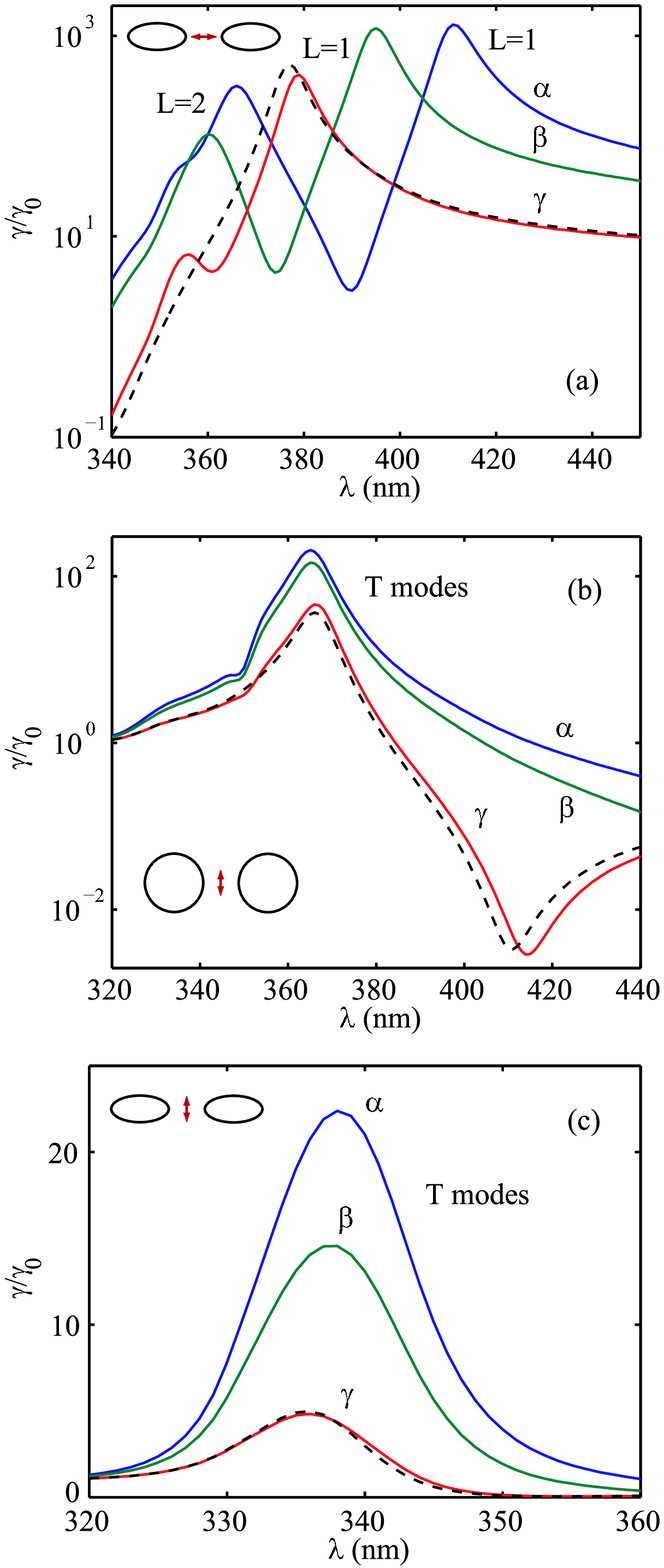}
\caption{\label{fig11} (Color online) Normalized radiative decay
rate of a dipole placed at the middle point between two identical
oblate nanospheroids made from silver as a function of the
wavelength. The dipole source moment is oriented along the x (a), y
(b), and z (c) axis. The large semi-axes of the nanospheroid are
$a=15$ nm, the aspect ratio is $c/a=0.6$. The curves $\alpha $,
$\beta $, $\gamma $ correspond to $l/2a = 1.05$, 1.1, and 1.3,
respectively. The asymptotic expression obtained by approximation of
the spheroids by point dipoles ($l/2a = 1.3$) is shown by the dashed
curve.}
\end{figure}

In Fig.~\ref{fig11}, normalized radiative decay rate of a dipole
placed at the middle point between the oblate nanospheroids is
shown. As one can see in Fig.~\ref{fig11}(a), if dipole moment of a
source is oriented along the line connecting the oblate
nanospheroids' centers, only plasmonic L-modes are excited as it
took place for the case of a cluster of two prolate nanospheroids
(cf. Fig.~\ref{fig10}(b)). When the distance between the
nanospheroids is large enough (the $\gamma$ curve), the properties
of radiative decay of a dipole source located near the cluster can
be more and more precisely approximated by means of the
self-consistent model where spheroids are modeled by point dipoles
(see Fig.~\ref{fig11}(a), the dashed curve).

From Fig.~\ref{fig11}(b) and (c) that correspond to the case of
dipole moment of a source oriented along the y and z axes, one can
conclude that only symmetric T-modes are presented here.
Nevertheless, at small enough distances between the nanospheroids
one can expect that M-modes are also excited but can not be seen due
to large losses (large imaginary part of dielectric permittivity) in
silver spheroids. If the imaginary part of permittivity of
spheroidal nanoparticles' material is small enough, for example, in
a case of silicized carbon (SiC) \cite{ref60}, all fine features of
the spectra and M-modes in particular will be well visible as it was
demonstrated in \cite{ref20,ref21} by the example of a cluster of
two spheres.

\section{\label{sect5}Conclusion}

Thus, in the present work optical properties of clusters made of two
metal nanospheroids are considered theoretically, and analytical
results are obtained. This investigation has become possible due to
proving of the addition translation theorem for spheroidal functions
in quasistatic regime. Plasmonic eigenoscillations were analyzed in
details, and it was found that in a cluster of two prolate or oblate
nanospheroids there can be three types of plasmon modes. Two of them
(low frequency, $0 < \omega < \omega _{pl} / \sqrt {2}$, L- and
T-modes) can be effectively excited by a plane electromagnetic wave,
while the third type (high frequency, $\omega _{pl} / \sqrt {2} <
\omega < \omega _{pl} $, M-modes) can be excited only by a strongly
nonuniform field of a nanolocalized source of light (a molecule, a
quantum dot) located in the gap between two adjacent nanoparticles.

We have also investigated excitation of a nano-antenna made from two
silver nanospheroids by fields of a plane wave and an electric
dipole. The results of these investigations allow us to find
absorption and scattering cross-sections of the nano-antenna as a
function of wavelength for various polarization of an incident plane
electromagnetic wave and to attribute all observable peaks to
excitation of corresponding plasmonic modes. We have also analyzed
radiative decay rate (or local density of state) of a dipole placed
in the gap between the nanospheroids and have attributed all
observable peaks to excitation of corresponding plasmonic modes.

The obtained analytical results can be used in many applications
based on plasmonic nano-antennas and on enhancement of local field
(SERS, SEF, nanolasers, nano-optical circuits, and so on). Besides,
our results are very important for controlling accuracy of different
computational software that has no a priori test of accuracy.

\begin{acknowledgments}
The authors thank Russian Foundation of Basic Researches (grant
09-02-13560) for financial support. The authors also express their
gratitude to Ulrike Woggon for her helpful comments.
\end{acknowledgments}

\appendix*
\section{\label{appen}Translational addition theorem for spheroidal wave functions in a quasistatic limit}

A solution of the Helmholtz equation in spheroidal coordinates can
be expanded as a series in spheroidal wave functions \cite{ref52}.
In spite of the fact that these functions are thoroughly studied
\cite{ref51,ref52,ref53,ref54,ref56} , the problem of
electromagnetic waves scattering on a spheroid still remains one of
the most complex ones, first of all, due to mathematical aspects
related to usage of spheroidal wave functions. On the other hand, it
is well-known that in a case of spheroids which size is
substantially less than the wavelength (quasistatic limit), a
solution of the Laplace equation in spheroidal coordinates can be
presented as a series in associated Legendre functions \cite{ref55}.
The Legendre functions are widespread and included as standard
special functions in majority of mathematical software programs,
such as {\textit{Maple}}, {\textit{Mathematica}}, {\textit{MatLAB}},
etc. That is why it is important to obtain the translational
addition theorem for spheroidal wave functions in the case when
sizes of spheroids are far less than the wavelength in which these
functions would be ``replaced'' by corresponding Legendre functions.

In \cite{ref45}, a rotational-translational addition theorem was
obtained for spheroidal wave functions in a case of an arbitrary
size of spheroids in comparison to the wavelength. This theorem was
derived in the following manner: first, spheroidal wave functions
were expanded over spherical wave functions, then the spherical wave
functions were expanded over spheroidal ones. Substituting one
expansion into the other and using the addition theorem for
spherical wave functions \cite{ref51,ref61}, the required theorem
for spheroidal functions was derived as a result. Here, we will do
the same in a particular case when spheroids' sizes are far less
than the wavelength.

In a case of a prolate spheroid, we have the following expansions
\cite{ref45}

\begin{eqnarray}
&&R_{mn}^{\left( {3} \right)} \left( {kf,\xi} \right)S_{mn}^{\left(
{1} \right)} \left( {kf,\eta} \right)e^{im\phi}  \nonumber \\
&&= {{\sum \limits_{q = {\left| {m} \right|},{\left| {m} \right|} +
1}^{\infty}} {' i^{q - n}d_{q - {\left| {m} \right|}}^{mn} \left(
{kf} \right)h_{q}^{\left( {1} \right)} \left( {kr} \right)P_{q}^{m}
\left( {\cos \theta} \right)e^{im\phi}} } , \label{eqa1}
\end{eqnarray}

\noindent where $1 \le \xi < \infty $, $ - 1 \le \eta \le 1$ and $0
\le \phi \le 2\pi $ are coordinates of the prolate spheroid
\cite{ref52}; $0 \le r < \infty $, $0 \le \theta \le \pi $ and $0
\le \phi \le 2\pi $ are spherical coordinates; $R_{mn}^{\left( {3}
\right)} \left( {kf,\xi} \right)$ and $S_{mn}^{\left( {1} \right)}
\left( {kf,\eta} \right)$ are radial and angular spheroidal
functions of the third and first kinds according to \cite{ref52};
$h_{q}^{\left( {1} \right)} \left( {kr} \right)$, and $P_{q}^{m}
\left( {\cos \theta} \right)$ are the spherical Hankel function of
the first kind \cite{ref56} and the associated Legendre function
\cite{ref56}; $d_{q - {\left| {m} \right|}}^{mn} \left( {kf}
\right)$ are the expansion coefficients \cite{ref52}; $k$ is
wavenumber, $f$ is a half of the spheroid's focal distance; the
stroke of the sum character implies summation over only even or only
odd values of $q$. Another important relation from \cite{ref45} has
the form:

\begin{eqnarray}
 &&j_{n} \left( {kr} \right)P_{n}^{m} \left( {\cos \theta}  \right)e^{im\phi}
= {\frac{{2}}{{2n + 1}}}{\frac{{\left( {n + m} \right)!}}{{\left( {n
- m}
\right)!}}} \nonumber \\
 &&\times {{\sum\limits_{q = {\left| {m} \right|},{\left| {m} \right|} +
1}^{\infty}} {{' \frac{{i^{q - n}}}{{N_{mq} \left( {kf}
\right)}}}d_{n - {\left| {m} \right|}}^{mq} \left( {kf}
\right)R_{mq}^{\left( {1} \right)} \left( {kf,\xi}
\right)S_{mq}^{\left( {1} \right)} \left( {kf,\eta}
\right)e^{im\phi}} } , \label{eqa2}
 \end{eqnarray}

\noindent where $j_{n} \left( {kr} \right)$ is the spherical Bessel
function \cite{ref56}; $R_{mq}^{\left( {1} \right)} \left( {kf,\xi}
\right)$ is a radial spheroidal function of the first kind
\cite{ref52}, and $N_{mq} \left( {kf} \right)$ is normalization
factor of angular spheroidal functions of the first kind
\cite{ref52}. In the case of a spheroid which size is far less than
the wavelength, that is, in the case $k \to 0$, one can use the
following asymptotic expressions \cite{ref51}:

\begin{eqnarray}
 d_{n - m - 2q}^{mn} &\approx& {\frac{{\left( {n + m} \right)!\left( {2n - 4q
+ 1} \right)!!\left( {2n - 2q - 1} \right)!!}}{{\left( {n + m - 2q}
\right)!\left( {2n + 1} \right)!!\left( {2q} \right)!!\left( {2n -
1} \right)!!}}}\left( {kf} \right)^{2q},\quad 0 \le q \le {\frac{{n
-
m}}{{2}}}, \nonumber \\
 d_{n - m + 2q}^{mn} &\approx& {\frac{{\left( { - 1} \right)^{q}\left( {n - m
+ 2q} \right)!\left( {2n - 1} \right)!!\left( {2n + 1}
\right)!!}}{{\left( {n - m} \right)!\left( {2n + 4q - 1}
\right)!!\left( {2q} \right)!!\left( {2n + 2q + 1}
\right)!!}}}\left( {kf} \right)^{2q},\quad q \ge 1, \nonumber
\\ \label{eqa3}
 \end{eqnarray}

\noindent and $N_{mn} \approx {\frac{{2}}{{2n + 1}}}{\frac{{\left(
{n + m} \right)!}}{{\left( {n - m} \right)!}}}$. Radial and angular
spheroidal functions $R_{mn}^{\left( {1} \right)} \left( {kf,\xi}
\right)$, $R_{mn}^{\left( {3} \right)} \left( {kf,\xi}  \right)$ and
$S_{mn}^{\left( {1} \right)} \left( {kf,\eta}  \right)$ can be also
presented in the asymptotic form \cite{ref51,ref53}

\begin{eqnarray}
 R_{mn}^{\left( {1} \right)} \left( {kf,\xi}  \right) &\approx& {\frac{{\left(
{n - m} \right)!}}{{\left( {2n - 1} \right)!!\left( {2n + 1}
\right)!!}}}\left( {kf} \right)^{n}P_{n}^{m} \left( {\xi}  \right), \nonumber \\
 R_{mn}^{\left( {3} \right)} \left( {kf,\xi}  \right) &\approx& -
i{\frac{{\left( { - 1} \right)^{m}\left( {2n - 1} \right)!!\left(
{2n + 1} \right)!!}}{{\left( {n + m} \right)!}}}\left( {kf}
\right)^{ - n - 1}Q_{n}^{m} \left( {\xi}  \right), \nonumber
\\ S_{mn}^{\left( {1} \right)} \left( {kf,\eta}  \right) &\approx&
P_{n}^{m} \left( {\eta}  \right), \label{eqa4}
 \end{eqnarray}

\noindent where $P_{n}^{m} \left( {\eta}  \right)$ is the associated
Legendre function \cite{ref56} defined on the segment $ - 1 \le \eta
\le 1$; $P_{n}^{m} \left( {\xi}  \right)$, $Q_{n}^{m} \left( {\xi}
\right)$ are associated Legendre functions of the first and second
kinds \cite{ref56}, correspondingly, defined in a complex plane with
branch cut from $ - \infty $ to +1. Finally, asymptotic forms for
spherical Bessel and Henkel functions have the form \cite{ref56}:

\begin{equation}
j_{n} \left( {kr} \right) \approx {\frac{{2^{n}n!}}{{\left( {2n + 1}
\right)!}}}\left( {kr} \right)^{n},\quad h_{n}^{\left( {1} \right)}
\left( {kr} \right) \approx - i{\frac{{\left( {2n}
\right)!}}{{2^{n}n!}}}\left( {kr} \right)^{ - n - 1}. \label{eqa5}
\end{equation}

Making use of Eqs.~(\ref{eqa3})-(\ref{eqa5}), one can obtain the
following relations from Eqs.~(\ref{eqa1}) and (\ref{eqa2}):

\begin{eqnarray}
 &&Q_{n}^{m} \left( {\xi}  \right)P_{n}^{m} \left( {\eta}  \right) =
{\sum\limits_{q = 0}^{\infty}  {{\frac{{\left( { - 1}
\right)^{m}\left( {n - m + 2q} \right)!\left( {n + m}
\right)!}}{{\left( {n - m} \right)!\left( {2n + 2q + 1}
\right)!!\left( {2q} \right)!!}}}\left( {{\frac{{f}}{{r}}}}
\right)^{n + 2q + 1}P_{n + 2q}^{m} \left( {\cos \theta}  \right)}} , \nonumber \\
 &&\left( {{\frac{{r}}{{f}}}} \right)^{n}P_{n}^{m} \left( {\cos \theta}
\right) = {\frac{{\left( {n - m} \right)!}}{{\left( {2n - 1}
\right)!!}}}P_{n}^{m} \left( {\xi}  \right)P_{n}^{m} \left( {\eta}
\right)
\nonumber \\
 &&+ {{\sum\limits_{q = 0,1}^{n - m - 2} } { ' {\frac{{\left( {n + m}
\right)!\left( {2m + 2q + 1} \right)q!}}{{\left( {n - m - q}
\right)!!\left( {n + m + q + 1} \right)!!\left( {2m + q}
\right)!}}}P_{m + q}^{m} \left( {\xi}  \right)P_{m + q}^{m} \left(
{\eta}  \right)}} . \label{eqa6}
 \end{eqnarray}

To obtain the required addition theorem for wave functions of
prolate spheroids which size is far less than the wavelength from
Eq.~(\ref{eqa6}), it is necessary to apply the corresponding theorem
for spherical wave functions \cite{ref19}:

\begin{eqnarray}
 &&\left( {{\frac{{l}}{{r_{j}}} }} \right)^{n + 1}P_{n}^{m} \left( {\cos
\theta _{j}}  \right)e^{im\phi _{j}}  = {\sum\limits_{q =
0}^{\infty} {{\sum\limits_{p = - q}^{q} {R_{pqmn} \left( {\theta
_{js} ,\phi _{js}} \right)\left( {{\frac{{r_{s}}} {{l}}}}
\right)^{q}P_{q}^{p} \left( {\cos
\theta _{s}}  \right)e^{ip\phi _{s}}} }} } , \nonumber \\
 &&R_{pqmn} \left( {\theta _{js} ,\phi _{js}}  \right) = {\frac{{\left( { - 1}
\right)^{q + m}\left( {q + n - p + m} \right)!}}{{\left( {q + p}
\right)!\left( {n - m} \right)!}}}P_{q + n}^{p - m} \left( {\cos
\theta _{js}}  \right)e^{i\left( {m - p} \right)\phi _{js}} .
\label{eqa7}
 \end{eqnarray}

\begin{figure}[there]
\includegraphics[width=8.0cm]{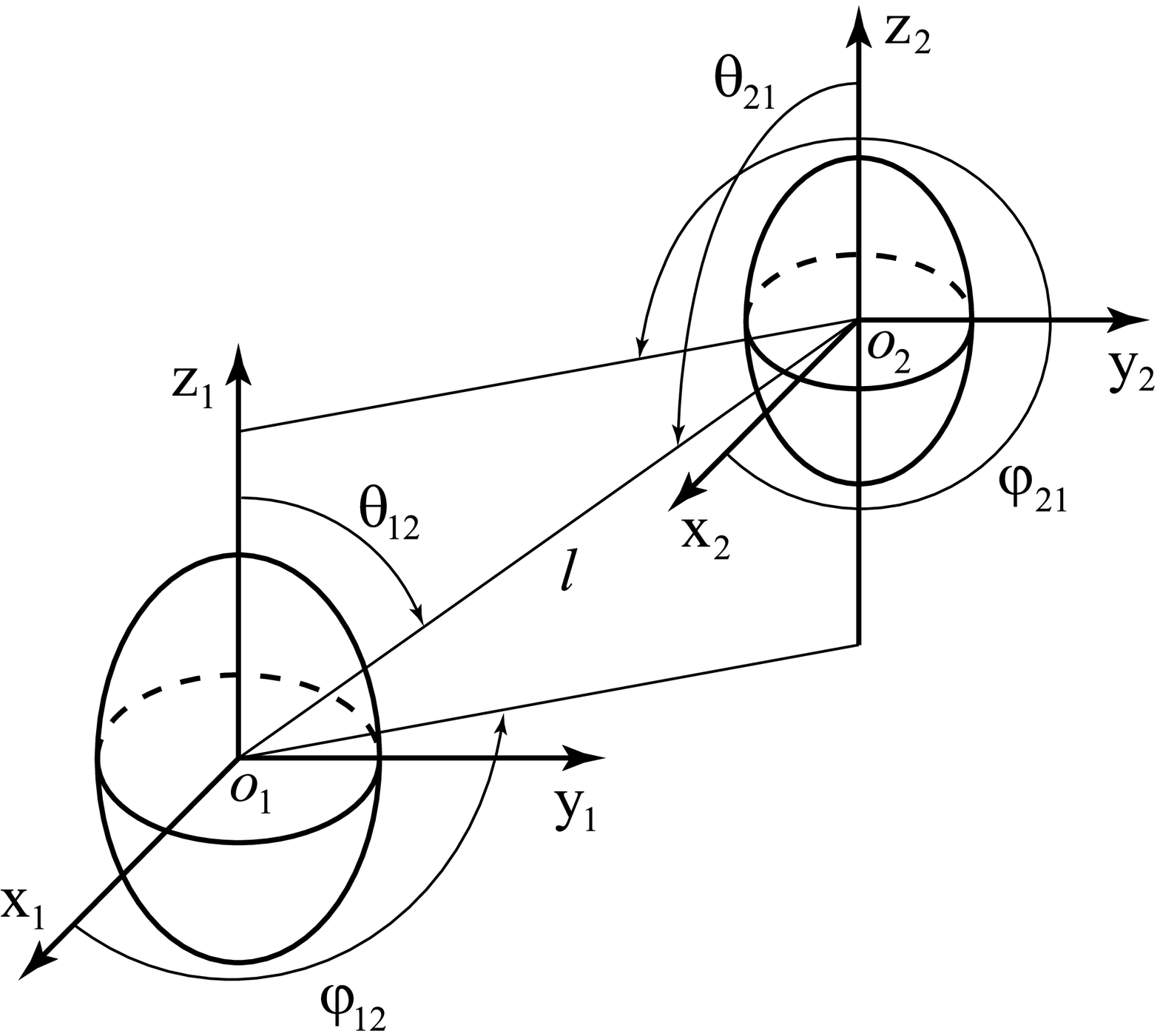}
\caption{\label{fig12} Geometry of derivation of the translational
addition theorem.}
\end{figure}

\noindent In the expression (\ref{eqa7}), the indices $j$, $s = 1$,
2 ($j \ne s$) denote the first ``1'' and the second ``2'' local
systems of spherical coordinates, that are based on local Cartesian
coordinates with parallel axes and attached to the spheres' centers;
($r_{js}$, $\theta _{js}$, $\phi _{js}$) are spherical coordinates
of the origin of the $s$-th coordinate system in the $j$-th local
system of coordinates; $l = r_{js} = r_{sj} $ is the distance
between the origins of local coordinates systems (see
Fig.~\ref{fig12}). Substituting Eq.~(\ref{eqa7}) into
Eq.~(\ref{eqa6}), we obtain after series of transformations the
required translational addition theorem for wave functions of
prolate spheroids in the quasistatic limit:

\begin{eqnarray}
 &&Q_{n}^{m} \left( {\xi _{j}}  \right)P_{n}^{m} \left( {\eta _{j}}
\right)e^{im\phi _{j}}  = {\sum\limits_{q = 0}^{\infty}
{{\sum\limits_{p = - q}^{q} {O_{pqmn} \left( {f_{j} ,f_{s} ,l,\theta
_{js} ,\phi _{js}} \right)P_{q}^{p} \left( {\xi _{s}}
\right)P_{q}^{p} \left( {\eta _{s}}
\right)e^{ip\phi _{s}}} }} } , \nonumber \\
 &&O_{pqmn} \left( {f_{j} ,f_{s} ,l,\theta _{js} ,\phi _{js}}  \right) =
{\frac{{\left( { - 1} \right)^{q}\left( {2q + 1} \right)\left( {q -
p} \right)!\left( {n + m} \right)!}}{{\left( {q + p} \right)!\left(
{n - m}
\right)!}}} \nonumber \\
 &&\times {\sum\limits_{r = 0}^{\infty}  {{\sum\limits_{k = 0}^{\infty}
{{\frac{{\left( {q + n + 2r + 2k - p + m} \right)!}}{{\left( {2q +
2r + 1} \right)!!\left( {2n + 2k + 1} \right)!!\left( {2r}
\right)!!\left( {2k}
\right)!!}}}}}} } \nonumber \\
 &&\times \left( {{\frac{{f_{s}}} {{l}}}} \right)^{2r + q}\left(
{{\frac{{f_{j}}} {{l}}}} \right)^{2k + n + 1}P_{q + n + 2r + 2k}^{p
- m} \left( {\cos \theta _{js}}  \right)e^{i\left( {m - p}
\right)\phi _{js}} . \label{eqa8}
 \end{eqnarray}

\noindent In Eq.~(\ref{eqa8}), the indices $j$, $s = 1$, 2 ($j \ne
s$) denote the first and second local prolate spheroid coordinate
systems with the parallel Cartesian axes introduced above; the
spherical angles $\theta _{js}$ and $\phi _{js}$ are the same as in
Eq.~(\ref{eqa7}); $l$ is the distance between the origins of local
coordinate systems (see Fig.~\ref{fig12} for details). Let us add
that an obligatory condition for usage of the theorem (\ref{eqa8})
is parallelism of Cartesian axes in the local coordinate systems. In
the case when the second local coordinate system is turned around
the first one, the theorem (\ref{eqa7}) is to be changed
\cite{ref19}, resulting in corresponding modification of the theorem
(\ref{eqa8}). Let us also note that mathematical applicability
condition for (\ref{eqa8}) is convergence of the series in this
expression. In particular, one may state that Eq.~(\ref{eqa8}) is
applicable in the case of non-overlapping spheroids \cite{ref51}.

The translational addition theorem for functions of oblate spheroids
which size is much less than the wavelength, can be obtained from
Eq.~(\ref{eqa8}) if the formal substitution is used \cite{ref52}:
$\xi \to i\xi$ and $f \to - if$. In expressions derived in this
manner, $\xi$ is a coordinate in the system of the oblate spheroid
($0 \le \xi < \infty )$; $f$ is a half of the focal distance of the
oblate spheroid. Let us note that (\ref{eqa8}) is a very general
theorem, and many other theorems relating prolate, oblate, and
spherical functions can be derived from it.

The obtained theorem (\ref{eqa8}) is not very convenient for
numerical calculations since it contains summation both over
positive and negative $p$. To simplify it, let us use properties of
the associated Legendre functions \cite{ref51,ref62}, basing on
which we shall find that

\begin{eqnarray}
 &&O_{ - p,q, - m,n} \left( {f_{j} ,f_{s} ,l,\theta _{js} ,\phi _{js}}
\right) \nonumber \\
 &&= \left( { - 1} \right)^{p - m}{\left[ {{\frac{{\left( {q + p}
\right)!\left( {n - m} \right)!}}{{\left( {q - p} \right)!\left( {n
+ m} \right)!}}}} \right]}^{2}O_{pqmn} \left( {f_{j} ,f_{s}
,l,\theta _{js} ,\phi _{js}}  \right)e^{i2\left( {p - m} \right)\phi
_{js}} . \label{eqa9}
 \end{eqnarray}

\noindent Combining Eqs.~(\ref{eqa8}) and (\ref{eqa9}), one can
derive the following relation ($m$, $p = 0$, 1, 2, \ldots ):

\begin{eqnarray}
 &&Q_{n}^{m}\left( {\xi _{j}}\right) P_{n}^{m}\left( {\eta _{j}}\right)
\left\{ {{\begin{array}{*{20}c} {\cos \left( {m\phi _{j}} \right)} \hfill \\
{\sin \left( {m\phi _{j}} \right)} \hfill \\ \end{array}}}\right. ={
\sum\limits_{q=0}^{\infty }{{\sum\limits_{p=0}^{q}{P_{q}^{p}\left(
{\xi _{s}}
\right) P_{q}^{p}\left( {\eta _{s}}\right) }}}}  \nonumber \\
&&\times {\left\{ {{\begin{array}{*{20}c} {L_{pqmn} \left( {f_{j}
,f_{s} ,l,\theta _{js}} \right)\cos \left( {p\phi _{s} - \left( {p -
m} \right)\phi _{js}} \right) } \hfill \\ {L_{pqmn} \left( {f_{j}
,f_{s} ,l,\theta _{js}} \right)\sin \left( {p\phi _{s} - \left( {p -
m} \right)\phi _{js}} \right) }
\hfill \\ \end{array}}}\right. }  \nonumber \\
&&\left. {{\begin{array}{*{20}c} { + T_{pqmn} \left( {f_{j} ,f_{s}
,l,\theta _{js}} \right)\cos \left( { - p\phi _{s} + \left( {p + m}
\right)\phi _{js}} \right)} \hfill \\ { + T_{pqmn} \left( {f_{j}
,f_{s} ,l,\theta _{js}} \right)\sin \left( { - p\phi _{s} + \left(
{p + m} \right)\phi _{js}} \right)} \hfill \\ \end{array}}}\right\}
,  \label{eqa10}
 \end{eqnarray}

\noindent where

\begin{eqnarray}
 &&L_{pqmn}\left( {f_{j},f_{s},l,\theta _{js}}\right) = {\frac{{\left( {-1}
\right) ^{q}\left( {2-\delta _{0p}}\right) \left( {2q+1}\right)
\left( {q-p} \right) !\left( {n+m}\right) !}}{{2\left( {q+p}\right)
!\left( {n-m}\right) !
}}} \nonumber \\
&&\times {\sum\limits_{r=0}^{\infty }{{\sum\limits_{k=0}^{\infty
}{{\frac{{ \left( {q+n+2r+2k-p+m}\right) !}}{{\left(
{2q+2r+1}\right) !!\left( {2n+2k+1}
\right) !!\left( {2r}\right) !!\left( {2k}\right) !!}}}}}}} \nonumber \\
&&\times {{{{\left( {{\frac{{f_{s}}}{{l}}}}\right) ^{2r+q}\left(
{{\frac{{ f_{j}}}{{l}}}}\right) ^{2k+n+1}P_{q+n+2r+2k}^{p-m}\left(
{\cos \theta _{js}}
\right) }}}}, \nonumber \\
&&T_{pqmn}\left( {f_{j},f_{s},l,\theta _{js}}\right) ={\frac{{\left(
{-1} \right) ^{q+m}\left( {2-\delta _{0p}}\right) \left(
{2q+1}\right) \left( {q-p }\right) !\left( {n+m}\right) !}}{{2\left(
{q+p}\right) !\left( {n-m}\right)
!}}} \nonumber \\
&&\times {\sum\limits_{r=0}^{\infty }{{\sum\limits_{k=0}^{\infty
}{{\frac{{ \left( {q+n+2r+2k-p-m}\right) !}}{{\left(
{2q+2r+1}\right) !!\left( {2n+2k+1}
\right) !!\left( {2r}\right) !!\left( {2k}\right) !!}}}}}}} \nonumber \\
&&\times {{{{\left( {{\frac{{f_{s}}}{{l}}}}\right) ^{2r+q}\left(
{{\frac{{ f_{j}}}{{l}}}}\right) ^{2k+n+1}P_{q+n+2r+2k}^{p+m}\left(
{\cos \theta _{js}} \right) }}}}, \label{eqa11}
 \end{eqnarray}

\noindent and $\delta _{0p} $ is Kronecker delta symbol.

In the case of coaxial prolate spheroids with a common axis $z,$ one
should take $\phi _{j} = \phi _{s} $ and $\theta _{js} = 0$ or
$\theta _{js} = \pi$ in Eq.~(\ref{eqa10}). This results in the
following relation ($m = 0$, 1, 2, \ldots ):

\begin{eqnarray}
 &&Q_{n}^{m} \left( {\xi _{j}}  \right)P_{n}^{m} \left( {\eta _{j}}  \right) =
{\sum\limits_{q = m}^{\infty}  {S_{mqmn} \left( {f_{j} ,f_{s}
,l,\theta _{js}}  \right)P_{q}^{m} \left( {\xi _{s}}
\right)P_{q}^{m} \left( {\eta
_{s}}  \right)}} , \nonumber \\
 &&S_{mqmn} \left( {f_{j} ,f_{s} ,l,\theta _{js}}  \right) = {\frac{{\left( {
- 1} \right)^{q}\left( {2q + 1} \right)\left( {q - m} \right)!\left(
{n + m} \right)!}}{{\left( {q + m} \right)!\left( {n - m}
\right)!}}}P_{q + n}
\left( {\cos \theta _{js}}  \right) \nonumber \\
 &&\times {\sum\limits_{r = 0}^{\infty}  {{\sum\limits_{k = 0}^{\infty}
{{\frac{{\left( {q + n + 2r + 2k} \right)!}}{{\left( {2q + 2r + 1}
\right)!!\left( {2n + 2k + 1} \right)!!\left( {2r} \right)!!\left(
{2k} \right)!!}}}\left( {{\frac{{f_{s}}} {{l}}}} \right)^{2r +
q}\left( {{\frac{{f_{j}}} {{l}}}} \right)^{2k + n + 1}}}} } .
\nonumber \\ \label{eqa12}
 \end{eqnarray}

In the case when local coordinate systems of the spheroids have a
common axis $x$, one should take $\theta _{js} = \pi / 2$ and $\phi
_{js} = 0$ or $\phi _{js} = \pi $ in Eq.~(\ref{eqa10}). As a result,
we have in this case

\begin{eqnarray}
&&Q_{n}^{m} \left( {\xi _{j}}  \right)P_{n}^{m} \left( {\eta _{j}}
\right){\left\{ {{\begin{array}{*{20}c}
 {\cos \left( {m\phi _{j}}  \right)} \hfill \\
 {\sin \left( {m\phi _{j}}  \right)} \hfill \\
\end{array}}}  \right.} \nonumber \\
 &&= {\sum\limits_{q = 0}^{\infty}  {{\sum\limits_{p =
0}^{q} {P_{q}^{p} \left( {\xi _{s}}  \right)P_{q}^{p} \left( {\eta
_{s}} \right){\left\{ {{\begin{array}{*{20}c}
 {M_{pqmn} \left( {f_{j} ,f_{s} ,l,\phi _{js}}  \right)\cos \left( {p\phi
_{s}}  \right)} \hfill \\
 {N_{pqmn} \left( {f_{j} ,f_{s} ,l,\phi _{js}}  \right)\sin \left( {p\phi
_{s}}  \right)} \hfill \\
\end{array}}}  \right.}}}} } , \label{eqa13}
\end{eqnarray}

\noindent where

\begin{eqnarray}
 &&M_{pqmn} \left( {f_{j} ,f_{s} ,l,\phi _{js}}  \right) \nonumber \\
 &&= \left( {L_{pqmn}
\left( {f_{j} ,f_{s} ,l,\pi / 2} \right) + T_{pqmn} \left( {f_{j}
,f_{s} ,l,\pi / 2} \right)} \right)\cos \left( {\left( {p - m}
\right)\phi _{js}}
\right), \nonumber \\
 &&N_{pqmn} \left( {f_{j} ,f_{s} ,l,\phi _{js}}  \right) \nonumber \\
 &&= \left( {L_{pqmn}
\left( {f_{j} ,f_{s} ,l,\pi / 2} \right) - T_{pqmn} \left( {f_{j}
,f_{s} ,l,\pi / 2} \right)} \right)\cos \left( {\left( {p - m}
\right)\phi _{js}} \right). \label{eqa14}
 \end{eqnarray}

\noindent To simplify Eq.~(\ref{eqa14}), it is convenient to use the
relation \cite{ref56}:

\begin{equation}
P_{n}^{m} \left( {0} \right) = {\frac{{2^{m}}}{{\sqrt {\pi}} } }\cos
\left( {\left( {n + m} \right){\frac{{\pi}} {{2}}}}
\right){\frac{{\Gamma \left( {\left( {n + m + 1} \right) / 2}
\right)}}{{\Gamma \left( {\left( {n - m + 2} \right) / 2}
\right)}}}, \label{eqa15}
\end{equation}

\noindent where $\Gamma \left( {x} \right)$ is Euler's gamma
function \cite{ref56}.

\pagebreak

\end{document}